\documentclass[%
reprint,
superscriptaddress,
 amsmath,amssymb, aps,
]{revtex4-2}

\usepackage{graphicx}
\usepackage{dcolumn}
\usepackage{bm}

\usepackage{amsmath}
\usepackage{amsfonts}
\usepackage{amssymb}
\usepackage{bm}
\usepackage{graphicx}
\usepackage[T1]{fontenc}
\usepackage{textcomp}
\usepackage{enumitem}
\usepackage{cancel}
\usepackage{bbold}
\usepackage{esvect}
\usepackage{commath}
\usepackage[most]{tcolorbox}
\usepackage{lipsum, babel}
\usepackage{comment}

\begin{document}

\title{Generation and control of non-local chiral currents in graphene superlattices\\ by orbital Hall effect}


\author{Juan Salvador-S\'{a}nchez}
\affiliation{%
 Nanotechnology Group, USAL—Nanolab, University of Salamanca\\
 Plaza de la Merced, Edificio Triling\"{u}e, 37008, Salamanca, Spain.
}%

\author{Luis M. Canonico}
\affiliation{%
 Catalan Institute of Nanoscience and Nanotechnology\\ CSIC and BIST, Campus UAB, Bellaterra, 08193 Barcelona, Spain.
}%

\author{Ana P\'{e}rez-Rodr\'{i}guez}%
\affiliation{%
 Nanotechnology Group, USAL—Nanolab, University of Salamanca\\
 Plaza de la Merced, Edificio Triling\"{u}e, 37008, Salamanca, Spain.
}%

\author{Tarik P. Cysne}
\affiliation{
Instituto de Física, Universidade Federal Fluminense, 24210-346 Niterói RJ, Brazil.
}%

\author{Yuriko Baba}
\affiliation{
GISC, Departamento de Física de Materiales, Universidad Complutense, 28040 Madrid, Spain.
}%

\author{Vito Cleric\`{o}}%
\affiliation{%
 Nanotechnology Group, USAL—Nanolab, University of Salamanca\\
 Plaza de la Merced, Edificio Triling\"{u}e, 37008, Salamanca, Spain.
}%

\author{Marc Vila}
\affiliation{%
 Catalan Institute of Nanoscience and Nanotechnology\\ CSIC and BIST, Campus UAB, Bellaterra, 08193 Barcelona, Spain.
}%
\affiliation{Department of Physics, University of California\\ Berkeley, California 94720, USA}
\affiliation{Materials Sciences Division, Lawrence Berkeley National Laboratory\\ Berkeley, California 94720, USA}

\author{Daniel Vaquero}%
\affiliation{%
 Nanotechnology Group, USAL—Nanolab, University of Salamanca\\
 Plaza de la Merced, Edificio Triling\"{u}e, 37008, Salamanca, Spain.
}%
\author{Juan Antonio Delgado-Notario}%
\affiliation{%
 CENTERA Laboratories, Institute of High Pressure Physics\\
 Polish Academy of Sciences, Warsaw 01-142, Poland.
}%
\author{Jos\'{e} M. Caridad}%
\affiliation{%
 Nanotechnology Group, USAL—Nanolab, University of Salamanca\\
 Plaza de la Merced, Edificio Triling\"{u}e, 37008, Salamanca, Spain.
}%
\author{Kenji Watanabe}
\affiliation{
Research Center for Functional Materials, National Institute for Materials Science\\
 1-1 Namiki, Tsukuba 305-0044, Japan.
}%

\author{Takashi Taniguchi}
\affiliation{
International Center for Materials Nanoarchitectonics,  National Institute for Materials Science\\
 1-1 Namiki, Tsukuba 305-0044, Japan.
}%

\author{Rafael A. Molina}
\affiliation{
Instituto de Estructura de la Materia, IEM-CSIC, E-28006 Madrid, Spain.
}%

\author{Francisco Domínguez-Adame}
\affiliation{
GISC, Departamento de Física de Materiales, Universidad Complutense, 28040 Madrid, Spain.
}%

\author{Stephan Roche}
\affiliation{%
 Catalan Institute of Nanoscience and Nanotechnology\\ CSIC and BIST, Campus UAB, Bellaterra, 08193 Barcelona, Spain.
}%
\affiliation{%
 ICREA–Institució Catalana de Recerca i Estudis Avançats, 08010 Barcelona, Spain
}%

\author{Enrique Diez}%
\affiliation{%
 Nanotechnology Group, USAL—Nanolab, University of Salamanca\\
 Plaza de la Merced, Edificio Triling\"{u}e, 37008, Salamanca, Spain.
}%

\author{Tatiana G. Rappoport}
 \email{tgrappoport@gmail.com}
\affiliation{Instituto de Telecomunicações, Instituto Superior Tecnico\\
University of Lisbon, Avenida Rovisco Pais 1, Lisboa, 1049001 Portugal
}
\affiliation{Instituto de Física, Universidade Federal do Rio de Janeiro\\
C.P. 68528, 21941-972 Rio de Janeiro RJ, Brazil}%

\author{Mario Amado}%
 \email{mario.amado@usal.es}
\affiliation{%
 Nanotechnology Group, USAL—Nanolab, University of Salamanca\\
 Plaza de la Merced, Edificio Triling\"{u}e, 37008, Salamanca, Spain.
}%

\date{\today}
             
\begin{abstract}
Graphene-based superlattices offer a new materials playground to exploit and control a higher number of electronic degrees of freedom, such as charge, spin, or valley for disruptive technologies. Recently, orbital effects, emerging in multivalley band structure lacking inversion symmetry, have been discussed as possible mechanisms for developing \emph{orbitronics}. Here, we report non-local transport measurements in small gap hBN/graphene/hBN moiré superlattices which reveal very strong magnetic field-induced chiral response which is stable up to room temperature. The measured sign dependence of the non-local signal with respect to the magnetic field orientation clearly indicates the manifestation of emerging orbital magnetic moments. The interpretation of experimental data is well supported by numerical simulations, and the reported phenomenon stands as a formidable way of in-situ manipulation of the transverse flow of orbital information, that could enable the design of orbitronic devices.
\end{abstract}



\maketitle


\section{\label{sec:level1}Introduction}
{ 
The electronic properties of graphene and other two-dimensional (2D) materials with a honeycomb lattice are dictated by the low-energy physics at two inequivalent $K$ and $K'$ valleys of the reciprocal space~\cite{CastroNeto2009}. The large separation between these valleys allows to distinguish valley quantum numbers that, likewise the spin degree of freedom, can be used to store and process information~\cite{Xiao07}. Moreover, the valleys in graphene possess opposite orbital magnetic moments, that at $K$ and $K'$ are proportional to the inverse of the band gap. Systems with small band gaps can have extremely large magnetic moments. This results in giant Zeeman splittings upon interaction with weak external magnetic fields, lifting valley degeneracy~\cite{Ge2021,Yin2022}.

In gapped graphene, the application of an electric field has been predicted to induce a flow of electrons, moving in opposite directions for different valleys, and thus giving rise to a valley Hall effect~(VHE)~\cite{Sui2015,Zhu2017}, that could be detected by non-local transport measurements~\cite{Gorbachev2014,Sui2015,Shimazaki2015,Zhu2017,Komatsu2018,Li2020}.  However, this is also accompanied by a transverse flow of valley magnetic moments. Consequently, the VHE can also be depicted as an orbital Hall effect \cite{Bhowal2021}. The orbital magnetic moments, which are physical quantities defined in the {\it entire momentum space}, replace the valley quantum numbers, that depend on the existence of well-defined pockets \cite{Bhowal2021}. Different from the VHE, this interpretation leads to a transverse current of a physical observable and can be used to analyze the interaction with magnetic and electric fields in the same framework.

Furthermore, electronic structure of graphene-based van der Waals (vdW2Ds) heterostructures can be tailored in a remarkable way by varying the twist angle between weakly interacting atomic layers and generating graphene moiré superlattices~\cite{Ponomarenko2013,Dean2013,Ribeiro2018,Gorbachev2014,Sui2015,Zhu2017,Komatsu2018,Li2020}. In particular, single or doubly aligned graphene/ hexagonal boron nitride (hBN) hetero\-junctions are very interesting for the study of inversion symmetry breaking in graphene. Indeed, such systems present considerable non-locality~\cite{Abanin2011,Gorbachev2014,Ribeiro2017,Komatsu2018,Aharon-Steinberg2021}, whose origin, although frequently associated to the VHE, is currently strongly debated~\cite{marmolejo2018deciphering,Renard2014,Aktor2021,Roche2022}. 
It has been shown that doubly aligned hBN/graphene stacks might generate a super-moiré pattern~\cite{Wang2019}, leading to the presence of small and non-uniform band gaps~\cite{Aktor2021,Roche2022}. vdW2Ds are, henceforth, a perfect platform for the study of inversion symmetry breaking in graphene. Because of the small gaps, the valley orbital magnetic moments are large and can be manipulated with magnetic and electric fields.

In this work, we report unambiguous formation of chiral non-local currents in doubly aligned hBN/graphene/hBN heterostructure, presenting direct evidence of their orbital magnetic origin. Our alignment design between layers minimizes the band gap, resulting in \textbf{giant orbital magnetic moments}. The interaction with weak magnetic fields lifts their degeneracy, generating chiral non-local currents. Quantum transport simulations in graphene nanoribbons with dispersive edge states and linear response theory calculations support our interpretation of the experimental findings regarding the origin of the chiral non-local resistance in graphene. 

\section{\label{sec:results}Results}

Figure~\ref{fig:device}\textbf{a} displays an optical image of the  hetero\-structure consisting of a graphite back gate ($15\,$nm thick) and monolayer graphene encapsulated between a top and bottom layer of hBN with a thickness of $10$ and $50\,$nm respectively. The crystals were aligned following their exfoliated straight edges using a micro-mechanical rotator and, for that reason, the relative twisting angles along the vertical heterostructure are expected to obey $m\times 30^\circ$, being $m = 0,\pm 1 ,\pm 2\dots$. From our electrical data displayed in Fig.~\ref{fig:mobility}\textbf{e} of the Supplementary Information we can reasonably discard twisting angles between graphene and hBN that are $0^\circ$ or a multiple of $60^\circ$. On the one side, charge neutrality point (CNP) appears as a standalone main Dirac peak with no traces of secondary satellite peaks in the measured carrier density range $\rvert n\rvert <3\times 10^{12}\,\mbox{cm}^{-2}$ that arise from the existence of electron-hole pockets at both sides of the main peak at such twisting angles~\cite{Abanin2011,Ponomarenko2013,Wang2015,Komatsu2018,Ribeiro2018,Finney2019,Li2020,Aharon-Steinberg2021}. On the other side, CNP resistivity of heterostructures where graphene is aligned  to $0^\circ$ or $60^\circ$ with the hBN exhibits a strong thermally activated behaviour with values exceeding hundreds of k$\Omega$ at low temperature, indicating a moiré coupling–induced band gap of the order of $30\,$meV~\cite{Abanin2011,Hunt2013,Wang2015,Ribeiro2018,Li2020}. In our case, we observe a thermally activated behaviour at low temperature but a CNP resistivity of only $\sim 7\,\mathrm{k}\Omega$ at room temperature. These characteristics are consistent with band gaps smaller than $10\,\mathrm{meV}$. Charge mobility extracted from magnetotransport Hall measurements rises to $200.000\,$cm$^2$/Vs, as shown in Fig.~\ref{fig:mobility} in the Supplementary Information. 

From a careful Raman analysis, we underpin the relative orientation of the flakes from the evolution of full-width at half-maximum of the 2D peak~(FWHM2D) as a function of the twisting angle. Finney \emph{et al.}~\cite{Finney2019} kept the bottom hBN aligned with graphene at $0^\circ$ and varied the relative angle of the top layer from $0^\circ$ to $60^\circ$. They showed that the vertical structure exhibits a noticeably high value of the FWHM2D, exceeding the standard one found on isolated graphene by 20 (40) cm$^{-1}$ if one (both) hBN layer(s) is (are) aligned to the graphene at the commensurate angles of $0^\circ$ or $60^\circ$~\cite{Finney2019}. Such broadening results from the moiré-scale relaxation of the graphene lattice, which strongly modifies the band structure~\cite{Eckmann2013}. We found a value of $\sim 20\,$cm$^{-1}$ for the FWHM2D  (see Fig.~\ref{fig:FWHM2D} of the Supplementary Information), in good agreement with the value for a standalone graphene flake that would correspond to a twisting angle that is neither $0^\circ$ nor $60^\circ$ in Ref.~\cite{Finney2019}. The ratio between the 2D and G peaks $I(2\mathrm{D})/I(\mathrm{G})>9$ also strengthens the assumption of twisting angles of $\pm 30^\circ$.}

Figure~\ref{fig:device}\textbf{c} shows the non-local resistances $R_{nl}$  for two different configurations (1 and 2) of the injection $i$ and collection $c$ terminals at $1.5\,$K. This allows us to infer the decay of the non-local signal as a function of the distance $\Delta x_i$ between the injection (local) and collection (non-local) terminals. For $\Delta x_1=2.5\,\mu$m, $R_{nl}\sim 1600\,\Omega$ which is consistent with other non-local measurements in graphene/hBN heterostructures~\cite{Abanin2011,Gorbachev2014,Li2020}. Moreover, the non-local signal gets weaker for increasing distances, reaching  $R_{nl}\sim 470\,\Omega$ for $\Delta x_2=5\,\mu$m. In the absence of external magnetic field, the position of the non-local peaks are centered around the CNP and are symmetric in respect to electron/hole regions. The relation $R_{nl}(\Delta x)=V_{nl}/I_0\simeq \pi\rho_{xx}e^{-\pi\rvert \Delta x\rvert/\text{W}}$, where $\text{W}=1.5\,\mu$m is the bar width, displays an exponential decay of the Ohmic contribution to $R_{nl}$ as a function of distance between the driving current and the non-local pair of contacts. This relation, already seen in graphene-based devices~\cite{Abanin2011,Ribeiro2017,Aharon-Steinberg2021}, serve us to rule out the Ohmic contribution to $R_{nl}$ as its prevailing mechanism. We extracted the ratio between the measured non-local resistances at different distances, obtaining ${R_{nl}(\Delta x_2)}/{R_{nl}(\Delta x_1)}\big\rvert_{\text{Measured}}=0.29$ while the purely-Ohmic expression gives us ${R_{nl}(\Delta x_2)}/{R_{nl}(\Delta x_1)}\big\rvert_{\text{Ohmic}} \sim 0.005$. From this analysis, one can see that  the non-local signal is orders of magnitude higher than the expected Ohmic contribution (see Fig.~\ref{exponential-decay} in the Supplementary Information).


\section{Chiral non-local signal}

\begin{figure*}
\includegraphics[width=15cm]{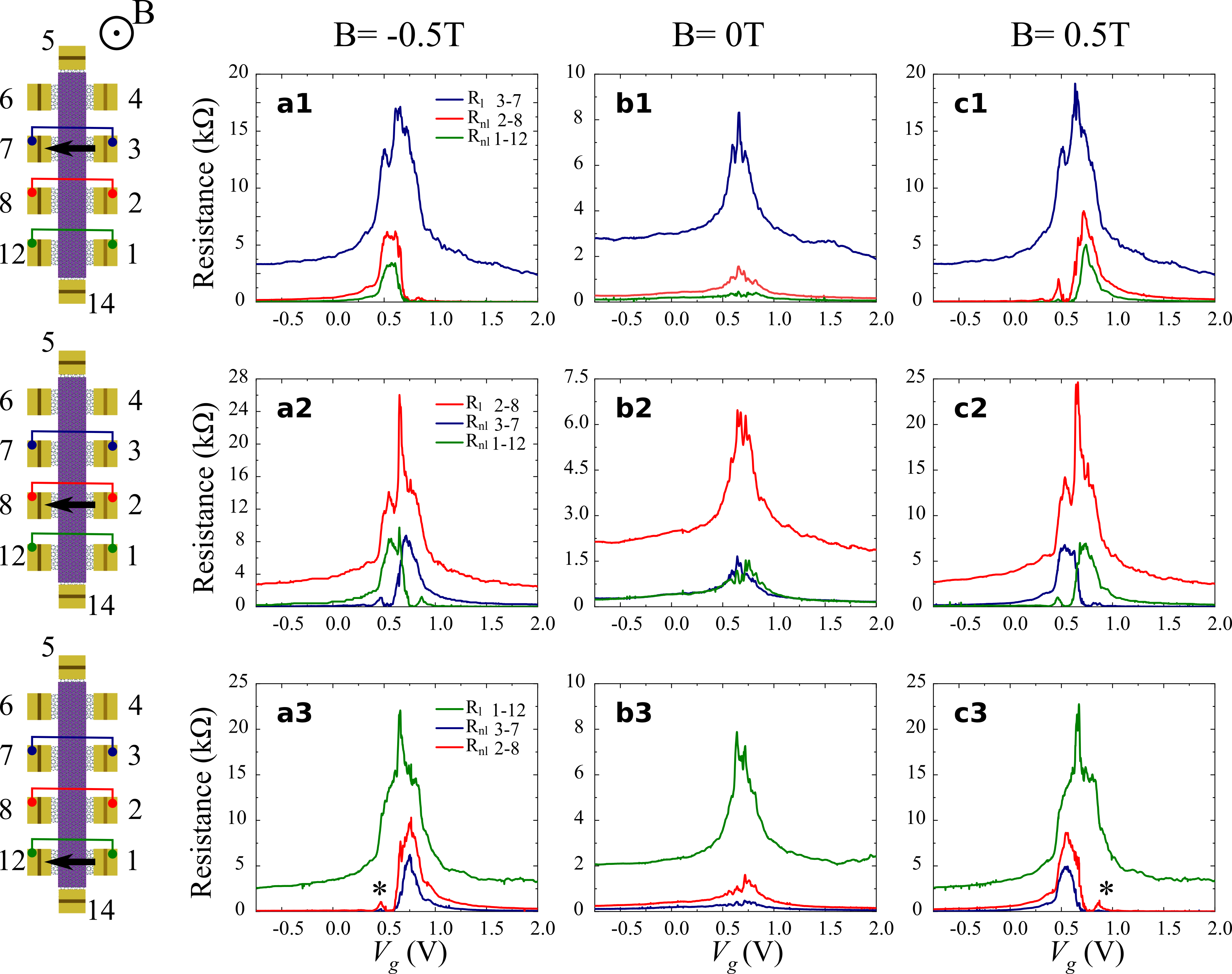}
\caption{\label{fig:9panels} Local (R$_{\text{l}}$) and non-local (R$_{\text{nl}}$) resistances for three different (1-3) configurations of the injection-collection terminals in the transversal direction as a function of the backgate voltage. The left column includes schematic top-view of the corresponding configuration for each row of graphs. Each column corresponds to a different value of the magnetic field, in the out-of-plane direction. The similarity of the non-local resistivity values in the absence of applied magnetic field shown in panel~\textbf{b2} evidences the homogeneity and, in general, the good quality of the sample. Moreover, similar values arise for opposing configurations (as in rows 1 and 3) for opposite directions of the applied magnetic field, as can be observed comparing panels~\textbf{a1} and \textbf{c3} with panels~\textbf{a3} and \textbf{c1}, for example.}
\end{figure*}

To explore the relation between the non-local currents and orbital magnetic moments, we apply positive and negative perpendicular magnetic fields and analyze the non-local signals for different injection-collection configurations. Unless stated otherwise, all measurements were performed at a fixed electronic temperature of $1.5\,$K and with an excitation current set to $10-20\,$nA. This low current amplitude was chosen to minimize thermal contributions to the non-local transport due to Joule heating and Ettingshausen effects~\cite{Renard2014} whilst simultaneously maximizing the signal-to-noise ratio of the measured voltages.  A summary of our results is shown in Fig.~\ref{fig:9panels}, which contains 9 different panels divided in three columns. Each column presents a different configuration of the external magnetic field, i.e., $-0.5\,$T, $0\,$T and $0.5\,$T.  Each of the three rows presents a different injection-collection setups, sketched in the diagram on the left. Each pair of contacts on the diagram has a specific color, while the arrows show the direction of the current for the pair of contacts of a particular injection-collection configuration. Each panel displays the local and non-local resistances as a function of the voltage applied to the graphite backgate. Their color palette matches the ones of the corresponding contacts, irrespective of the specific injection-collection setup. 

Let us first comment on the effect of the magnetic field on the local resistances.  The CNP is located at $V_g\sim0.6\,$V, as extracted from the Lorentzian fit of the local resistance, and does not present a noticeable shift when the magnetic field is applied. Still, there is a sharp increase of the local resistance with the magnetic field, which is very pronounced at the CNP.
Panel~\textbf{b1} in Fig.~\ref{fig:9panels} displays the evolution of the non-local signal as a function of the distance without external magnetic field. Panel~\textbf{b2} represents a configuration of the non-local pairs of contacts that are placed symmetrically to the current flow but at opposite directions.  In this case, the homogeneity of the sample is demonstrated because the non-local signals at both sides of the current flow have an expected matching value, as the magnitude of the non-local signal decays with the absolute value of the distance to the injection current. This characteristic, discussed previously, can also be seen in panels~\textbf{a2} and~\textbf{c2} of Fig.~\ref{fig:9panels}. 

A striking behavior of the non-local signal arises in presence of an external magnetic. We first focus on the case where the current is injected between two non-local contacts. Figure~\ref{fig:9panels}.\textbf{a2} shows the non-local resistances for the $B=-0.5\,$T, where one can see a clear separation between the peaks of opposite contacts. Moreover, they are mostly located either in the electron or hole sectors. Surprisingly, the position of the two peaks is swapped upon magnetic field reversal, which is a clear indication of a chiral behavior of the electronic response. If the two collectors are located at the same side of the injector, as shown in the first and last row of Fig.~\ref{fig:9panels}, the situation is different. In these two cases, the two non-local resistance peaks are centered and located either at the electron or hole sector and switch positions with the sign of the magnetic field and the relative orientation with respect to the collector. While both peaks appear at the hole sector in Fig.~\ref{fig:9panels}.\textbf{a1}, when the sign of $B$ is reversed, they appear at the electron sector (see Fig.~\ref{fig:9panels}.\textbf{c1}). If instead, we switch the position for positive $B$, as in Fig.~\ref{fig:9panels}.\textbf{a3}, the peaks also appear at the electron sector, changing to the hole sector if the field is reversed (see Fig.~\ref{fig:9panels}.\textbf{c3}). Furthermore, it is important to mention, that in resemblance to similar experiments, the non-locality is strongly enhanced with the magnetic field in all configurations~\cite{Abanin2011,Ponomarenko2013,Aharon-Steinberg2021}.

\begin{figure*}
\includegraphics[width=1.0 \linewidth]{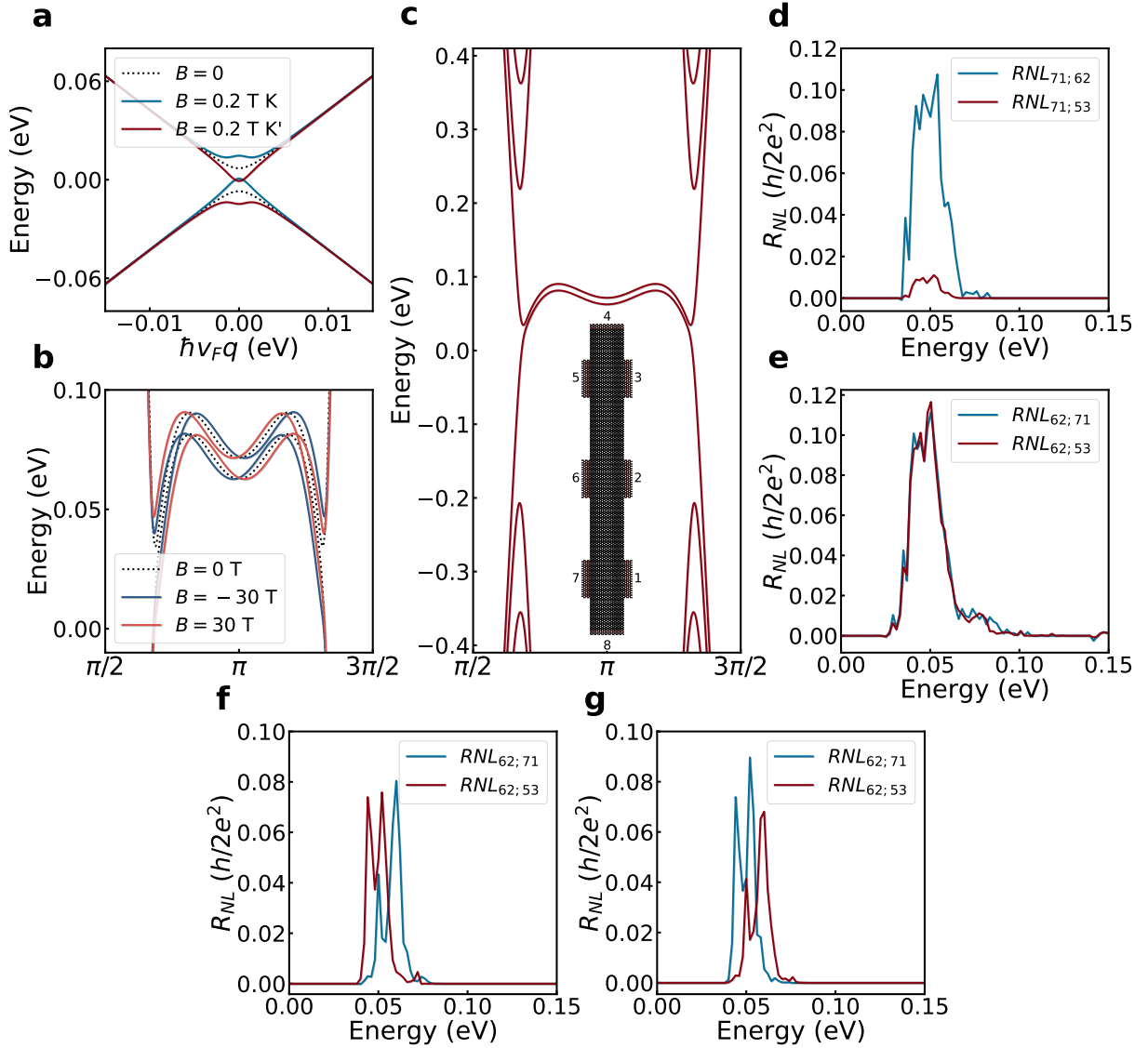}
\caption{\label{fig:nlfig} \textbf{a}, Comparison between the energy bands of gapped graphene systems with and without out-of-plane magnetic fields with $B=0.2\,$T and a gap $\Delta = 14\,$meV. \textbf{b}, Comparison of the energy bands of zigzag nanoribbons in for $B=0$ and $|B|=30\,$T and sublattice staggered potential $V_{ab}=5\,$meV. \textbf{c}, Energy bands of zigzag gaped graphene nanoribbons with $V_{ab}=5\,$meV. The inset shows the schematic representation of the device used in the simulations. Non-local resistance for current injection across terminals $0-5$ (panel \textbf{d}) and $1-4$ (panel \textbf{e}) in the absence of external magnetic fields. Non-local resistance for current injected across terminals $1-4$ for $B=-30\,$T (panel \textbf{f}) and $B=30\,$T (panel \textbf{g}).}
\end{figure*}

The nine panels in Fig.~\ref{fig:9panels} demonstrate a fully chiral behaviour of the non-local signal at low magnetic fields, which has not been observed in similar heterostructures in monolayer graphene with twisting angle set to different commensurate angles \cite{Abanin2011,Li2020,Aharon-Steinberg2021}. To clarify the underlying mechanism, we use the modern theory of magnetism and numerical simulations based on the linear response theory and the Landauer-B\"uttiker formalism implemented in the \textsc{KWANT} toolkit~\cite{groth2014kwant}. Addressing the orbital magnetic moment in solids is a nontrivial topic due to the ill-defined behaviour of the $\widehat{\bm r}$ operator in the Bloch basis~\cite{OrbitalMagnetizationVanderbitlt,xiao2010berry,MappingTopologicalOrderRealSpaceResta}. Nonetheless, modern theory of magnetism provides an appropriate description of this phenomenon by treating the Bloch electron as a self-rotating wave packet whose magnetic moment is expressed purely in bulk quantities as ${\bm m_{n}({\bm k})} = -i(e/2\hbar) \langle\nabla_{\bm k}U_n|\times\left[ H({\bm k}) - \varepsilon_{n}({\bm k})]\right]|\nabla_{\bm k}U_n\rangle$, where $n$ is the band index and $U$ is the periodic part of the Bloch eigenstate. Following Ref.~\cite{cai2013magnetic}, one can show that applying this approach to gapped Dirac materials leads to an expression for the orbital magnetic moment that reads ${\bm m({\bm k})} = (\tau e\hbar/2m^*)\left(1 +  v_F^2k^2/\Delta ^{2} \right)^{-1}\,{\bm \hat{ z}}$, where $\tau=\pm 1$ is the valley quantum number, $m^*=\hbar^2\Delta/v_F^2$ is the effective mass at the Dirac point (DP), $\Delta$ is the system gap and $v_F$ is the Fermi velocity. In this case, the intensity of the orbital magnetic moment is inversely proportional to the system gap while it can generally be seen as inversely proportional to the effective mass as well. This orbital moment in the presence of weak magnetic fields behaves similarly to the spins, i.e. it can couple directly to them. This gives rise to a $k$-dependent Zeeman effect that, in first order perturbation theory, renormalizes the energy spectrum close to the Dirac points~\cite{cai2013magnetic,zhou2019valley}, as shown in Fig.~\ref{fig:nlfig}\textbf{a}. It can also be interpreted in the usual way as a Zeeman effect with giant g-factor, which is proportional to the inverse of the effective mass, a well known effect in semiconductors. The valleys have opposite magnetic moments, producing a relative shift between the valleys. This effect can be very strong for small gap systems, leading, for example, to situations where the Fermi energy lies inside the gap for one valley while it is located in the electron/hole sector for the other.

Given the proportionality with the inverse of the gap size and coupling with external fields, one can argue that the chiral behavior observed in the non-local resistance measurements appears as a manifestation of the \textbf{orbital Zeeman effect} allowed by the formation of small a gap in the doubly-encapsulated hBN/graphene/hBN heterostructure. Considering the OHE formalism, one should expect a chiral behavior as observed in Fig.~\ref{fig:9panels} (see the Supplementary Information for details). The magnetic moments of each valley flow in opposite directions. Because of the sign of the Berry curvature, the flows invert direction when switching from hole to electron sectors. At the same time, the Zeeman shift between the valleys leads to a relative shift between the non-local peaks  that should invert with the change in the magnetic field orientation and collector's locations. Using this reasoning, we can estimate the band gap from non-local peaks of Fig.~\ref{fig:9panels}.\textbf{a2}, obtaining $\Delta\sim 5-8\,$meV. 

To inquire about this, we calculated the non-local resistance in an eight terminal device containing a graphene nanoribbon with the geometry shown in the inset of panel~\textbf{c} of Fig.~\ref{fig:nlfig} in the presence of external magnetic fields. The advantage of this approach is that it does not rely on any assumption regarding the orbital Zeeman effect, as the magnetic field is taken into account by the well established Peierls' substitution. Moreover, it can directly connect with experiments, as it allows the calculation of the non-local resistances. Finally, it considers both bulk and edge states. Still, here we are not particularly interested in the controversy related to the specific channels conveying the non-local current because, as it will become clear below, the orbital Zeeman effect is present in bulk and edge states and either of them can, in principle, produce chiral non-local currents. 

Previous numerical simulations showed the need of dispersive edge states near the Dirac point for non-local transport in gapped graphene~\cite{marmolejo2018deciphering}. They are absent in theories based on the simplistic Hamiltonian considering a single $p_z$ orbital. Instead, we used a $6$--bands tight-binding model that takes into account also the $d$ orbitals~\cite{boykin2011accurate} and a staggered sublattice potential to break the inversion symmetry of the system. It is important to note that other effects such as non-uniform potential and coupling to hBN layers can also lead to dispersive edge states and similar non-local currents (see Sec.~7 of the Supplementary Information for more details).

Panels~\textbf{b} and~\textbf{c} of Fig.~\ref{fig:nlfig} show the energy bands for a zig-zag graphene nanoribbon with $13\,$nm of width and a sublattice staggered potential of $5\,$meV. From panel~\textbf{c}, it is noticeable that the energy bands from this multi-orbital model exhibit a dispersive behavior similar to the ones observed from \emph{ab-initio\/} calculations of Ref.~\cite{marmolejo2018deciphering} without a gap in the whole spectrum, but have valence-conduction band separation due to the inversion symmetry breaking and very well-defined electron pockets at opposite sides in the nanoribbon Brillouin zone. Panel \textbf{b} depicts the comparison of the energy bands of the nanoribbon in the cases without considering magnetic fields and with out-of-plane magnetic fields via Peierls' substitution. The most relevant feature displayed by this panel is the stark valley-contrasting coupling with the magnetic field, which is similar to the one observed for the bulk bands. Although the modern theory of magnetism is well-developed only for bulk systems, it is clear, from the results in this panel, that the behavior displayed by the energy bands of the nanoribbon is in qualitative agreement with this theory. 

Aiming to reproduce the results from the middle columns in Fig.~\ref{fig:9panels}, in panels~\textbf{d} and~\textbf{e} we used the same injection-collection scheme for our non-local resistance simulations. Panel~\textbf{d} portrays the case in which the injection occurs at one side of the device. As in the measurements, the simulations results show considerable decay of the non-local signal with the channel length. On the other hand, panel~\textbf{e} shows the case in which the current injection occurs at the middle terminals of the device and the non-local resistance is computed in the leads of its sides. Comparing the non-local signal in the lateral terminals, it is clear that the non-local signal at both sides of the sample is overall symmetric. However, the most striking behavior appears when opposite magnetic fields are taken into account, as in the cases displayed in panels~\textbf{f} and~\textbf{g}. Comparing these figures with the experimental measurements shown in Fig.~\ref{fig:9panels}, we find convincing agreement. Although, due to the small sizes of the simulation as compared to the experiment, we need to apply much higher magnetic fields to obtain comparable results. Nevertheless, the most outstanding feature of these results becomes evident when comparing them with the renormalized energy bands in panel~\textbf{b}, where the chirality and energy selectivity is directly related to the coupling between the orbital magnetic moment and the external magnetic field, suggesting that the mechanism at play in the generation of these non-local signals observed in the experiment is the orbital valley Hall effect~\cite{Bhowal2021}. Still, our analysis suggests that Fermi surface edge currents carry the non-local signal, once the absence of dispersive edge states destroys the non-local signal.

\begin{figure*}
\includegraphics[width=1.0\linewidth]{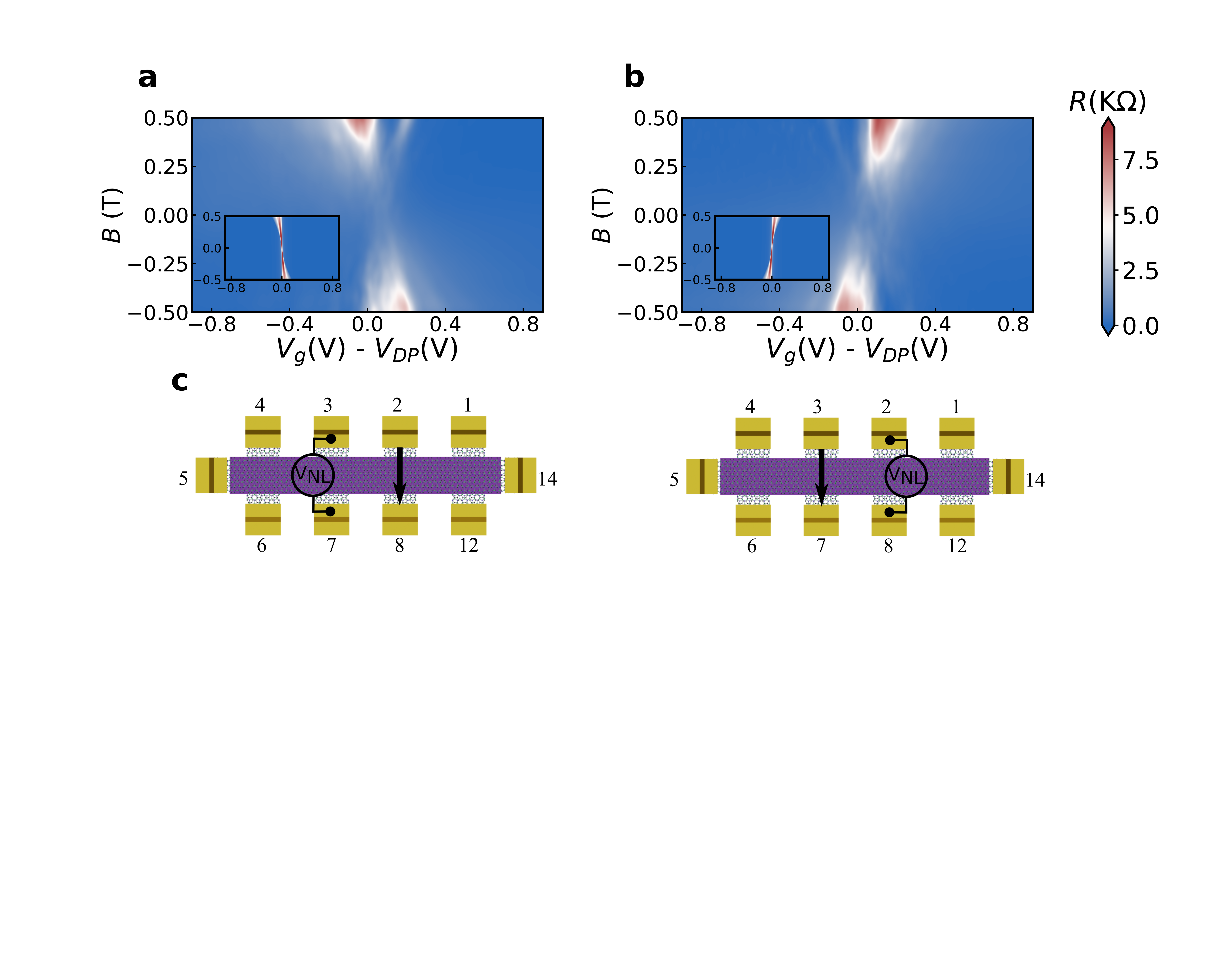}
\caption{Panels~\label{fig:chiral} and~\textbf{a} and \textbf{b} show the heat maps of the sample non-local resistance for two symmetrical configurations scanned in the space of backgate voltage, corrected by the position of the Dirac peak at $B=0$ ($V_{G} - V_{DP})$, and magnetic field measured at $250\,$mK. Insets on each graph show the numerical results for the OHC using Eq.~(\ref{sigmaOH}) in the Supplementary Information, showing extraordinary agreement with the experimental results. \textbf{c}, Sketch for the two different injection and collection configurations for the experimental and theoretical results shown in panels~\textbf{a} and~\textbf{b} above.} 
\end{figure*}

Figures~\ref{fig:chiral}\textbf{a} and~\ref{fig:chiral}~\textbf{b} display the contour plot of $R_{nl}$ as a function of both $B$ and $V_g-V_{DP}$ for two symmetrical configurations for the local and non-local contacts as sketched below each panel. The electronic temperature at which the curves were recorded was $T=250\,$mK and the voltage has been centered at the Dirac peak for $B=0$. These two panels show a clear chiral behavior and an apparent valley-carrier locking in the non-local signal for low magnetic fields  ranging from $-0.5$ to $0.5\,T$. In Fig.~\ref{fig:chiral}\textbf{a} we can observe a distinct transition from a hole-mediated non-local transport for negative magnetic fields towards an electron-like one when the magnetic field is reversed. Moreover, a strong asymmetry in the non-local curves is clearly visible, with a sudden decay while approaching the DP from the dominant carrier species towards the prohibited one.  Figure~\ref{fig:chiral}\textbf{b} displays the mirrored configuration for the pairs of contacts, compared to Fig.~\ref{fig:chiral}\textbf{a} as sketched in panel Fig.~\ref{fig:chiral}\textbf{c}. Consequently, we can argue that the valley-carrier locking, visible in the non-local signal, is a quite robust phenomenon and implies a flow of carriers characterized by different orbital magnetic moments. As the non-local signal should originate from the flow of orbital magnetic moments from a single valley, we compare $R_{nl}$ with the valley filtered OHE, calculated according to Ref.~\cite{Bhowal2021}, shown in the insets of  Figs.~\ref{fig:chiral}\textbf{a} and~\ref{fig:chiral}\textbf{b}. In conclusion, the interpretation of the experimental results are very consistent with the changes in the OHE resulting from the orbital Zeeman effect. The theoretical calculations indicate that the observed Zeeman shift is consistent with a graphene gap of $\Delta\sim 5-10\,$ meV. For details on the OHE calculations, see Sec.~6 of the Supplementary Information. 

Figure~\ref{fig:Zeeman} presents the local and non-local signals for a configuration where the excitation current lies symmetrically between two different pairs of contacts for $R_{nl}$ at a fixed perpendicular magnetic field of $B_\perp=0.5\,$T and variable in-plane component ($B_\parallel$) measured at $1.5\,$K. $B_\parallel$ ranges from $0$ to $12\,$T and its evolution has been marked with an arrow as a guide to the eye. When no parallel magnetic field is applied ($B_\parallel=0$), the charge carrier type is effectively coupled to one of the valleys and the asymmetric chiral behavior is found. Distinctively, while enhancing the in-plane magnetic field component, both non-local and local signals diminish and become symmetric. From these data, it becomes clear that the chiral non-local currents cannot be attributed to a spin-dependent effect. In-plane magnetic fields can be used to probe spin-polarized currents, as they lead to a non-local resistance that oscillates in function of the field. At the same time, they can also be used as a tuning parameter to manipulate the orbital characteristics of 2D multilayers~\cite{Roy2013,Kwan2020}. They affect the quasi-momentum of each layer differently~\cite{Pershoguba2010, Asakawa2017,Denner2020,Wei2021}, modifying the effective coupling between the layers and thus the resulting band-structure. Therefore, similar to strain, they can modify substantially the physics of twisted bilayers and moiré superlattices~\cite{Roy2013,Kwan2020}. 

Still, the application of an in-plane magnetic field $B_{\parallel}$ introduces a layer-dependent gauge field $\mathbf{A}_l =\mathbf{B}_{\parallel}\times \mathbf{z}_l$ that modifies the electron momentum $\mathbf{p}\rightarrow \mathbf{p} +(e/c)\mathbf{A}_l$, where $l$ indexes the layer. For graphene encapsulated by two hBN layers and located at $z=0$, the magnetic field shifts the momenta of electrons in each hBN layer along opposite directions. Time-reversal symmetry is broken, shifting the momenta of the electrons of different valleys in the same direction. This can have an important impact in the resulting band-structure, modifying the orbital magnetic moments and henceforth modifying local and non-local signals.

In conclusion, we have presented non-local transport measurements on hBN/graphene/hBN narrow gap heterostructures at low magnetic fields, which clearly indicate the presence of chiral effects. Such chiral response is inferred from the non-local resistance when reversing both the magnetic field and the injection-collection configurations. The interaction between large orbital magnetic moments arising in small gap graphene based superlattices and external magnetic field produces a relative Zeeman shift between the two valleys in both bulk and edge electron states. Furthermore, based on our experimental and theoretical analysis, it is clear that, regardless of the details about the location of current flow, the manifestation of strong chiral effect originates from the interplay between the Zeeman shift and the transverse flow of orbital magnetic moments. Importantly, the analysis of the non-local transport as a function of the magnetic field direction rules out spin effects, whereas its dependence with the distance between contacts clarifies that Ohmic and thermal contributions bring marginal contributions to the main signal. Finally, our computational results show that the valley-orbital Hall effect displays fingerprints in both bulk and edge transport, being of topological nature or not. All these facts support the interpretation that the origin of the giant non-locality in the studied graphene superlattices is linked to the orbital Hall effect resulting from the valley magnetic moments. Beyond shining a new light on a fierce debate concerning the formation of topological versus non-topological valley-driven phenomena to explain previously reported non-local signals \cite{Torres_2021, Roche_2022}, our findings pave the way towards future developments in room-temperature graphene orbitronics. 

\begin{acknowledgments}

M.~A. and E.~D. acknowledge financial support from the Ministerio de Ciencia, Innovaci\'{o}n y Universidades of Spain (Spanish Ministry of Science, Innovation, and Universities) and FEDER (ERDF: European Regional Development Fund) under the Research Grants numbers PID2019-106820RB-C21/22, PGC2018-094180-B-I00 and FEDER/Junta de Castilla y León Research Grant number SA121P20 D.~V. acknowledges financial support from the Ministerio de Universidades (Spain) (PhD contract FPU19/04224), including funding from ERDF/FEDER. J.~S.-S. acknowledges financial support from the Consejer\'{\i}a de Educaci\'{o}n, Junta de Castilla y Le\'{o}n, and ERDF/FEDER. L.~M.~C. acknowledges funding from Ministerio de Ciencia e Innovaci\'{o}n de Espa\~{n}a under grant No. PID2019-106684GB-I00 / AEI /10.13039/501100011033. S.~R. acknowledges funding from the European Union Seventh Framework Programme under grant no. 881603 (Graphene Flagship). The Catalan Institute of Nanoscience and Nanotechnology (ICN2) is funded by the CERCA Programme/Generalitat de Catalunya and supported by the Severo Ochoa programme (MINECO grant no. SEV-2017-0706). T.~G.~R. acknowledges funding from Fundação para a Ciência e a Tecnologia and Instituto de Telecomunicações - grant number UID/50008/2020 in the framework of the project Sym-Break. J.~M.~C. acknowledges support from the MICINN Ram\'{o}n y Cajal program (Project No. RYC2019-028443-I). M.~V. was supported as part of the Center for Novel Pathways to Quantum Coherence in Materials, an Energy Frontier Research Center funded by the U.S. Department of Energy, Office of Science, Basic Energy Sciences. K.~W. and T.~T. acknowledge support from JSPS KAKENHI (Grant Numbers 19H05790, 20H00354 and 21H05233). J.~A.~D.-N. acknowledges funding from by the CENTERA Laboratories in the framework of the International Research Agendas program for the Foundation for Polish Sciences, co-financed by the European Union under the European Regional Development Fund (No. MAB/2018/9). L.~M.~C. acknowledges funding from the project “Conversiópn de energía en heteroestructuras de van der Waals” funded by the Agencia Estatal de Investigación (AEI) (PID2019-106684GB-I00/ AEI/10.13039/501100011033)

\end{acknowledgments}

\textbf{Author contributions:} : M.~A., E.~D. and F.~D.-A. developed the concept of the experiment. T.~T. and K.~W. provided hBN crystals. J.~S.-S., A.~P.-R., V.~C., D.~V., and J.~A.~D.-N. performed device fabrication and carried out Raman spectroscopy. J.~S.-S., A.~P.-R. E.~D., and M.~A. performed transport measurements. J.~S.-S. and A.~P.-R. performed experimental analysis. J.~S.-S., A.~P.-R., E.~D. and M.~A. interpreted results with help from L.~M.~C., T.~G.~R., S.~R., J.~M.~C., Y.~B., R.~A.~M and F.~D.-A. Theoretical calculations were performed by L.~M.~C., T.~G.~R., T.~P.~C., M.~V. and S.~R. The manuscript and the Supplementary Materials were written by T.~G.~R., M.~A., S.~R., L.~M.~C., J.~S.-S., A.~P.-R. and F.~D.-A. with additional contributions from all authors.

\begin{figure*}
\includegraphics[width=15cm]{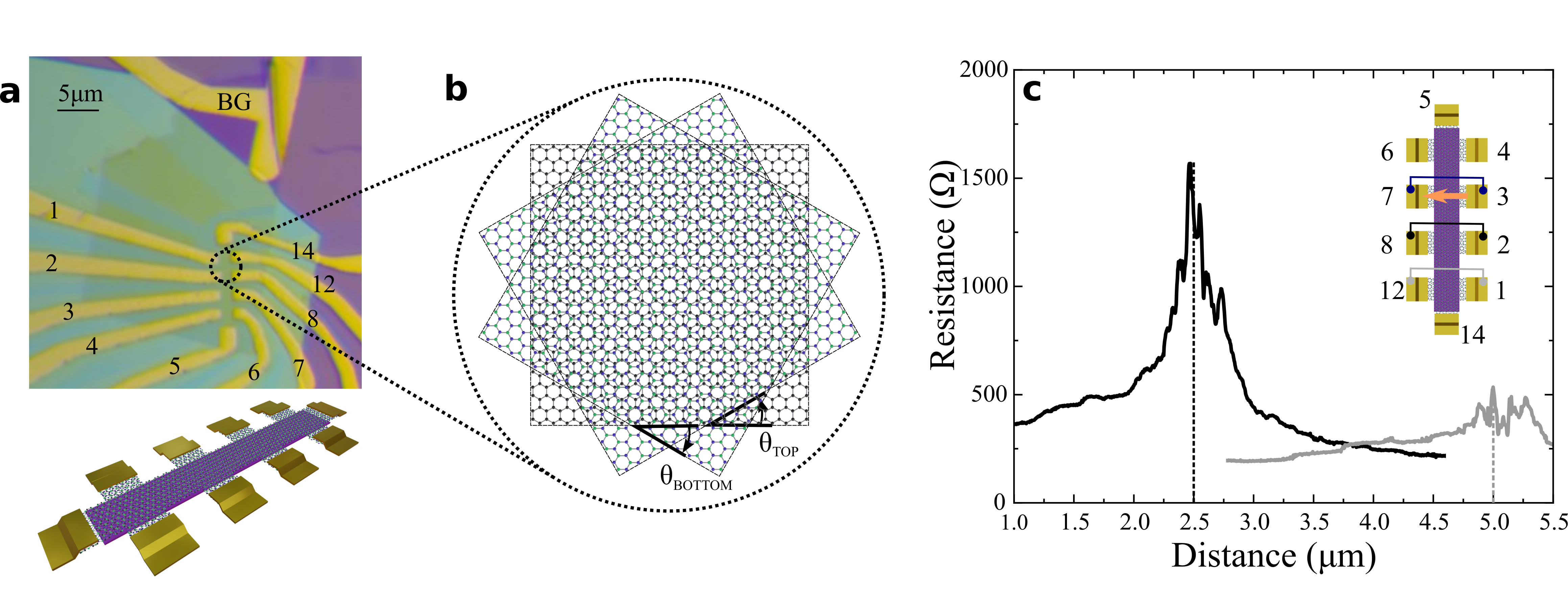}
\caption{\label{fig:device} \textbf{a}, Optical image and schematics of a complete device, consisting on a heterostructure based on a graphite back gate ($15\,$nm) and monolayer graphene encapsulated between two layers of hBN with thicknesses $10\,$nm (top) and $50\,$nm (bottom). \textbf{b}, Art view schematic of the lattice configuration with twist angles ($\theta_{\text{\tiny TOP}}$ and $\theta_{\text{\tiny BOTTOM}}$) between graphene and the top/bottom h-BN, assuming $\theta_{\text{\tiny TOP}}$  $\approx 30^{\circ}$. A moiré wavelength of $\lambda\sim 0.47\,$nm can be extracted from the relation found in Refs.~\cite{Yankowitz2012,Li2020}, in stark contrast with the usual $\lambda\sim14\,$nm present in fully-aligned samples. \textbf{c}, Non local resistance dependence with the distance to the injection source for two different sets of contacts at $T=1.5\,$K in absence of external magnetic field. The two pairs of contacts are separated $2.5$ and $5.0\,\mu$m from the local signal. The solid line serve as the direction of the driving (local) current. Dashed lines represent the position of the maximum of the two non-local resistances as a guide to the eye.}
\end{figure*}

\textbf{Methods:} The device fabrication of the superlattices follows the standard dry transfer technique with a polycarbonate film fabricated and deposited onto a polydimethylsiloxane stamp. The relative rotation between the different layers, following their natural edges, was controlled using a heated stage with a micromenchanical rotator with an accuracy better than $0.5^\circ$. The heterostructure rested atop a commercial Si/SiO$_{2}$ substrate. The fabricated stack was patterned using electron beam lithography followed by a dry-etching process in an ICP-RIE to define the sample geometry. The sample was patterned into a form of a multiterminal Hall bar (optical image in Fig. \ref{fig:device}) with the central horizontal bar of width W = 1.5 $\mu$m, total length of $\sim11\,\mu$m and a distance between the centres of the contacts of $2.5\,\mu$m. Electrical contact to all devices was made by Cr/Au (10nm/50nm) deposited by electron beam evaporation. Supplementary Sections 1 Figure \ref{fig:S1} and \ref{fig:S2} provide full details of the device fabrication.  We extracted the all Raman spectra and their associated FWHM2D from a Lorentzian fit (see Supplementary Section 2 Figures \ref{fig:Raman} and \ref{fig:FWHM2D}). Transport measurements in the multiterminal device were conducted in two- and four-terminal geometries with a.c. current excitation of $10$-$20\,$nA using the standard lock-in technique at 17.7Hz. The graphite layer was gated by applying a direct bias to it in the range of $\pm 12\,$V.

\textbf{Data availability:} All data needed to evaluate the findings of this study are present in the paper and/or the Supplementary Information. Data are available from the corresponding authors on reasonable request.

\textbf{Competing interests:} The authors declare that they have no competing interests.




\bibliography{references}

\begin{figure*}
\includegraphics[width=15cm]{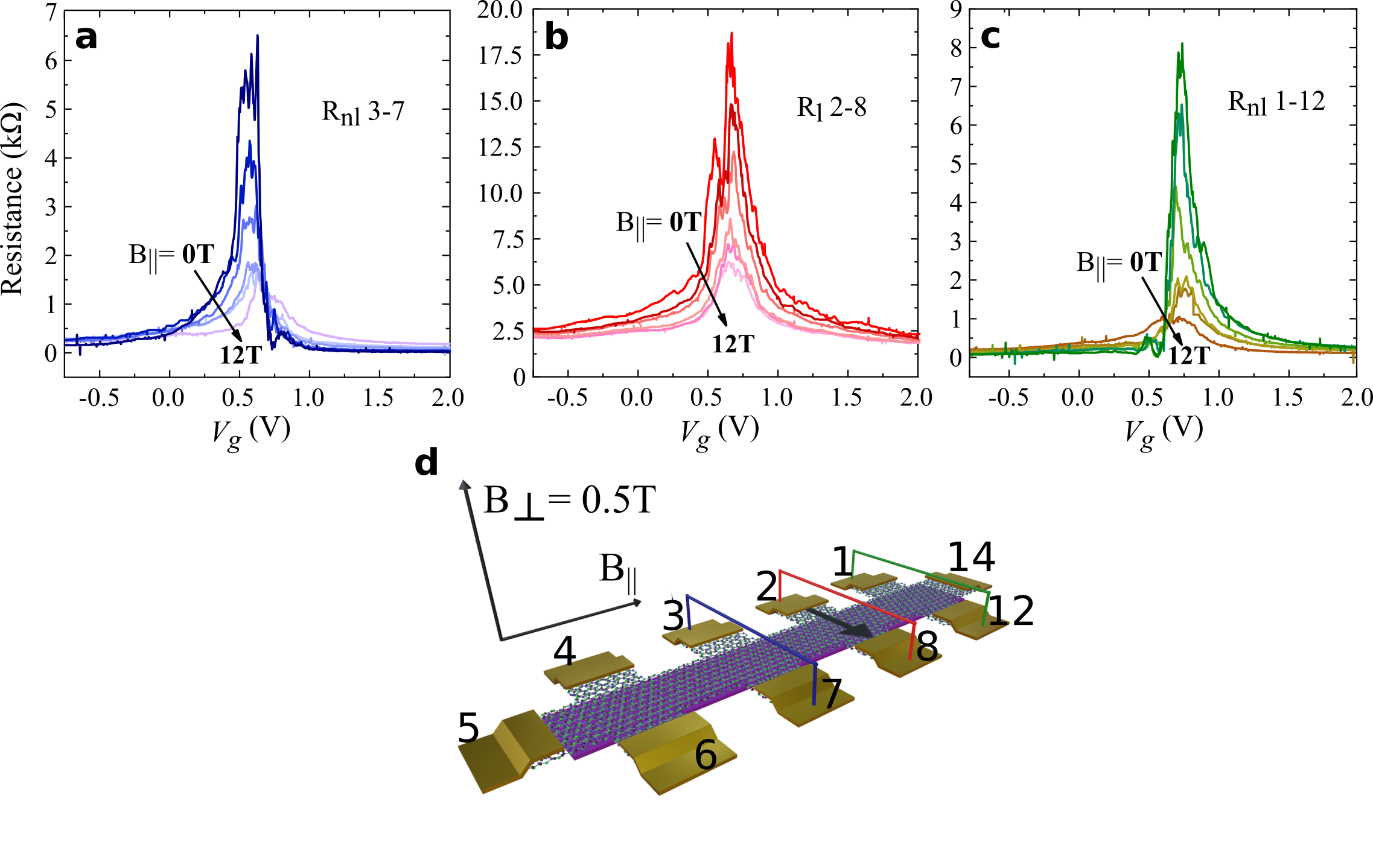}
\caption{\label{fig:Zeeman} Evolution of the non-local (\textbf{a} and \textbf{c}) and local (\textbf{b}) resistances as a function of the external in-plane magnetic field for a fixed out-of-plane field $B_{\perp}= 0.5\,$T. Panel~\textbf{d} sketched the sample, showing the direction of the driving current, the two pairs of contacts for non-local measurements and the two components for the external magnetic field. Solid arrows show the evolution of the in-plane component for $B$ ranging from $0$ to $12\,$T.
}
\end{figure*}

\clearpage
\begin{widetext}
\appendix
\section{Supplementary Information}
\subsection{Sample fabrication of the graphene-based heterostructures.}

 Mechanical exfoliation using an adhesive tape was performed on natural crystals of hexagonal boron nitride (hBN) and graphite on a silicon oxide wafer (290nm). h-BN flakes were used as top (Fig.~\ref{fig:S1}\textbf{a}) and bottom (Fig.~\ref{fig:S1}\textbf{b}) dielectric layers, as well as a 50nm graphite flake (Fig.~\ref{fig:S1}\textbf{d}). The thickness was confirmed through measurements in  a Stylus Profilometer. For the stacking process of the heterostructure, a polycarbonate (PC) film was fabricated and deposited on polydimethylsiloxane (PDMS). The top hBN flake was picked up at 50-60$^{\circ}$ and deposited on the graphene monolayer at 190$^{\circ}$, using the natural edges of the different layers to align the crystalline directions of the two materials. Afterwards,the hBN/Graphene heterostructure was cleaned to remove the polycarbonate film by rinsing in chloroform for a few minutes. The same technique was used to deposit the hBN bottom flake onto the graphite back gate. Finally, the stacked hBN top and graphene flakes were picked up in a similar manner and deposited onto the hBN bottom and graphite stack and aligned with natural edges.\\
 
 The fabricated stack was patterned using electron beam lithography (EBL-SEM). Firstly, a 320nm thick layer of PMMA in chlorobenzene 4$\%$ (by weight) was spin coated onto the stack. EBL-SEM was employed to define the sample geometry (Fig.~\ref{fig:S2}-panel a), followed by a dry-etching process in an ICP-RIE Plasma Pro Cobra 100 with a SF$_6$ atmosphere (40 sccm, P = 6 mTorr, P = 75 W at 10 ºC)  ((Fig.~\ref{fig:S2}-panel b). A second electron-beam lithography step (employing PMMA in chlorobenzene 5$\%$) and ICP-RIE Plasma-etching process was done in a similar manner for defining the the side contacts area (Fig.~\ref{fig:S2}-panels c and d) \cite{Delgado2022}. The contacts were deposited via evaporation of Cr/Au (10nm/50nm) followed by the standard liftoff procedure. The final device, forming a Hall bar with a central horizontal bar of W=1.5$\mu$m, a total length of 11$\mu$m, and a distance between the contacts of $L = 2.5\,\mu$m as shown in panel e. The device was bonded on a LCC20 chip carrier for electrical characterisation.
 
\begin{figure*}[!h]
\includegraphics[width=15cm]{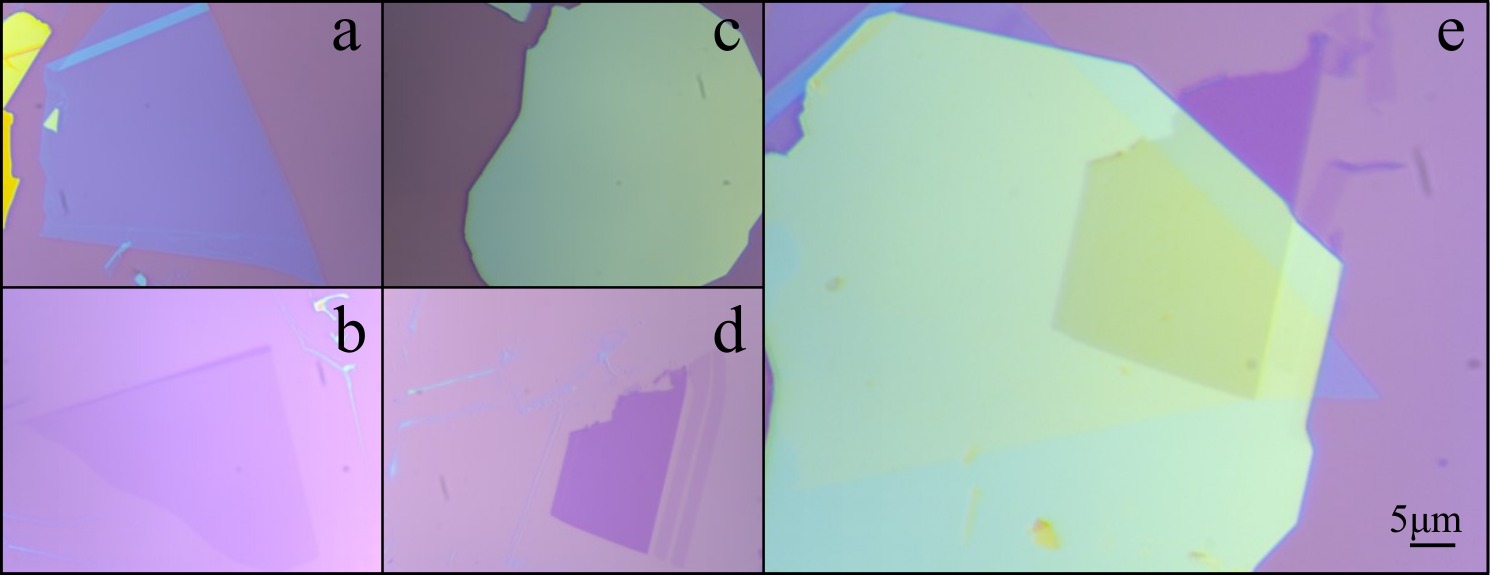}
\caption{\label{fig:S1} SUPPORTING INFO 1. Optical images of the different vdW2Ds layers involved in the fabrication of the sample through the stacking process. \textbf{a}, 10-nm-thick top hBN. \textbf{b}, Single layer graphene. \textbf{c}, 50-nm-thick bottom hBN. \textbf{d}, Graphite back gate. Thickness for the different flakes was confirmed through Raman and profilometer measurements. \textbf{e}, Resulting final vertical heterostructure.} 
\end{figure*}

\begin{figure*}
\includegraphics[width=15cm]{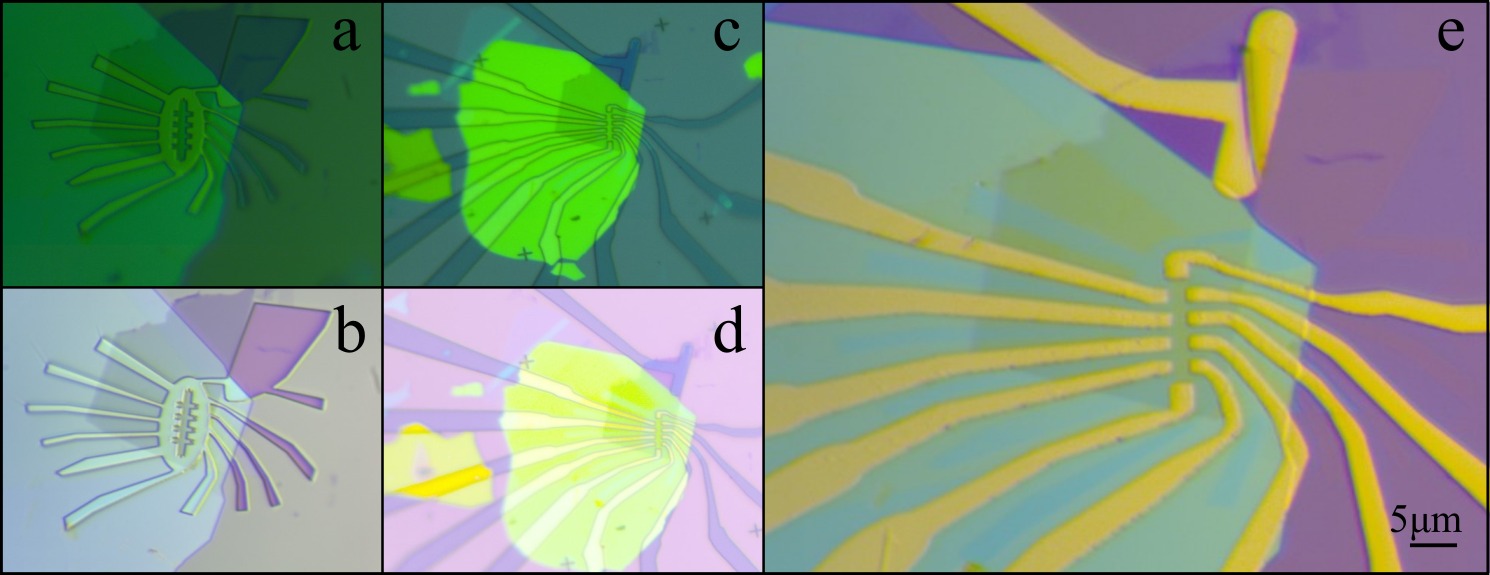}
\caption{\label{fig:S2} SUPPORTING INFO 2. Optical images of the sample through the Hall bar fabrication process via EBL lithography. The resulting devices is displayed in panel e.}
\end{figure*}

\clearpage
\subsection{Raman characterization.}

Each flake and the final heterostructure were characterized prior to their stacking by Raman spectroscopy using a micro-Raman spectrometer LabRAM HR Evolution at a wavelength 532 nm and an incident power of $\sim$10 mW. Figure \ref{fig:Raman} presents a detailed analysis of the Raman spectra found in the final heterostructure. Panel a) displays an optical image of the final stack prior to the electron beam lithography with a complete 2D Raman map in panel b) obtained at XXX cm$^{-1}$). Panels c) to e) show individual Raman spectra taken at different positions of the stack where only encapsulated graphene was present \textbf{c}, graphite back-gate is found \textbf{d} and the whole final structure where the Hall bar was designed \textbf{e}. 

From a careful Raman analysis we can study the relative orientation of the flakes conforming the stack as shown in Ref.~\cite{Finney2019} where the evolution of the value of the full-width at half-maximum of the 2D peak (FWHM2D) as a function of the twisting angle is presented. In that work, Finney \emph{et al.} maintained the bottom hBN aligned to the graphene at $0^\circ$ and varied the relative angle of the top one ranging from $0^\circ$ and $60^\circ$ showing that the vertical structure exhibits a noticeably broadened value of the FWHM2D exceeding the standard one found on isolated graphene by 20 (40) cm$^{-1}$ if one (both) hBN layer(s) is (are) aligned to the graphene at the commensurate angles of $0^\circ$ or $60^\circ$. Such broadening result from the moiré-scale relaxation of the graphene lattice which will strongly modify the graphene band structure~\cite{Eckmann2013}. In our case, a value of $\sim 20\,$cm$^{-1}$ for the FWHM2D is found (Figure. S5 Figure.~\ref{fig:FWHM2D}), in good agreement with the value obtained for a standalone graphene flake that would correspond to a twisting angle that is neither $0^\circ$ nor $60^\circ$ in Ref.~\cite{Finney2019}. 

\begin{figure*}[!h]
\includegraphics[width=10cm]{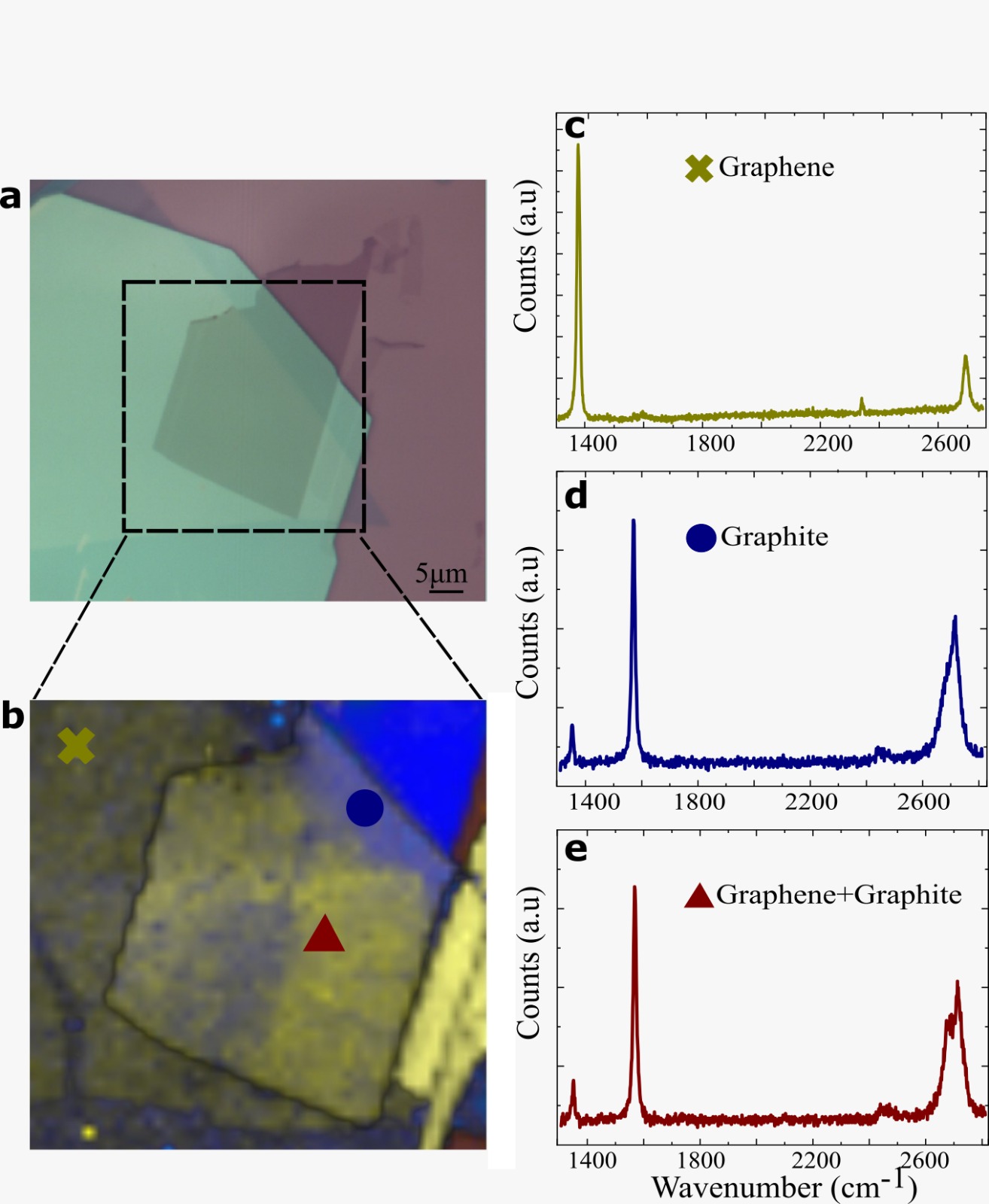}
\caption{\label{fig:Raman}SUPPORTING INFO 3. \textbf{a}, Optical image of the vertical heterostructure. \textbf{b}, Raman spectra of the whole designated area. \textbf{c},\textbf{d},\textbf{e}, Full Raman spectra of the positions marked with the coloured symbols in (\textbf{b}) where encapsulated graphene is present (\textbf{c}), the graphite back gate is found (\textbf{d}) and both graphene and graphite coincide in the underlying structure (\textbf{e}).}
\end{figure*}
\begin{figure*}
\includegraphics[width=15cm]{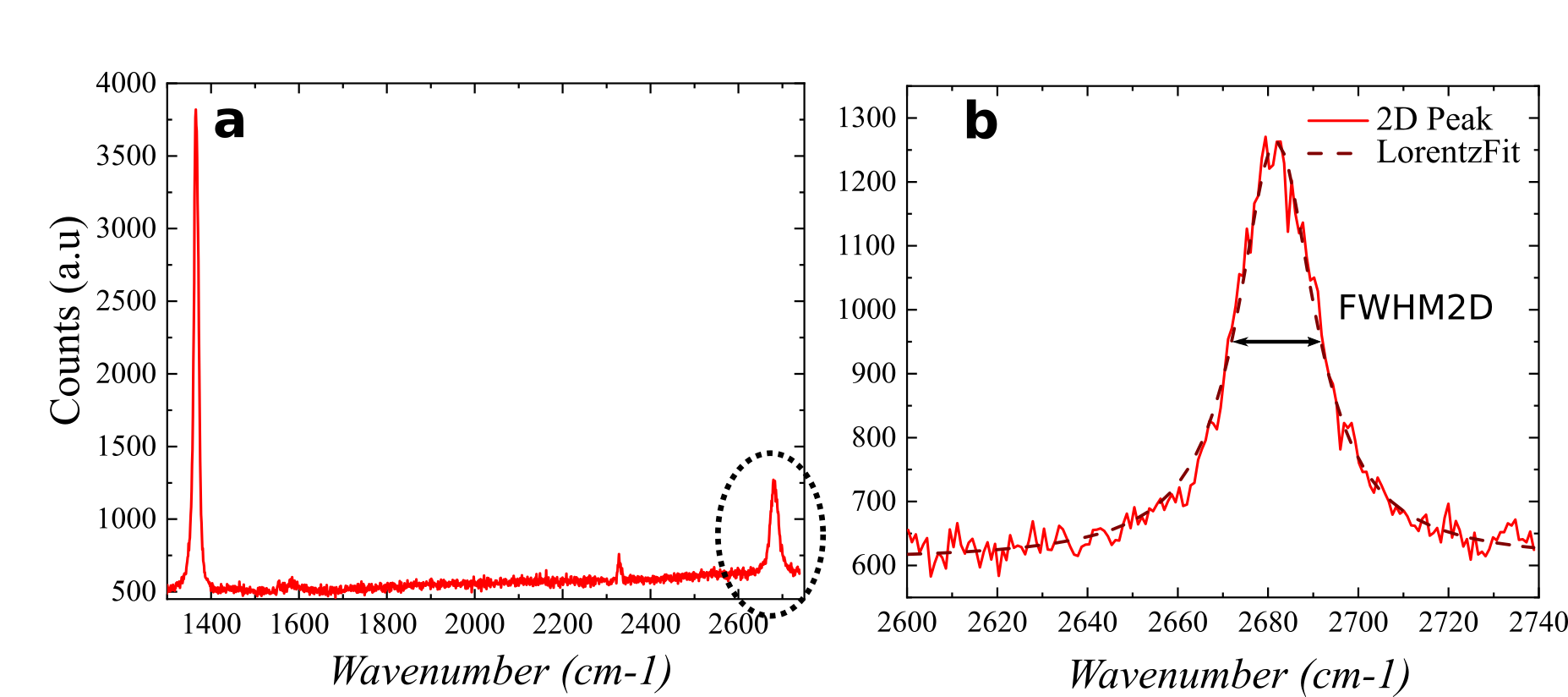}
\caption{\label{fig:FWHM2D}SUPPORTING INFO 4. \textbf{a}, Raman spectrum of the encapsulated single layer graphene flake. \textbf{b}, 2D peak extracted data from the highlighted region in (\textbf{a},) and its Lorentzian fit. A value for the full-width at half-maximum of the 2D peak of FWHM2D$\sim 20\,$cm$^{-1}$ is found which will rule out twisting angles between the layers in the vicinity of $\theta_{\text{\tiny TOP}}, \theta_{\text{\tiny BOTTOM}} \sim 0^\circ, \sim \pm 60^\circ$ as shown in Ref.~\cite{Finney2019}}
\end{figure*}

\clearpage
\subsection{Magnetotransport at low temperatures. Quantum Hall effect and mobility.}

Longitudinal ($R_{\text{XX}}$) and Hall ($R_{\text{XY}}$) resistances as a function of the external magnetic field are shown in figure \ref{fig:mobility}. Every spectrum was measured at $1.5\,$K at different values of the back gate voltage, namely $0\,$V, $0.3\,$V and $1.5\,$V from panel \textbf{a} to \textbf{c}. Values for the mobility have been extracted from the well-known formula from magnetotransport measurements where $\mu=1/(R_\square\cdot q\cdot n)$ where $n$ is the carrier density, $q$ the charge of the carrier and $R_\square=R_{\text{XX}(0)}\frac{W}{L}$ at a every given value of the backgate voltage. Results for the mobility for both holes and electrons consistently lie in the range of $\mu \sim 200,000$cm$^2/$Vs, comparable to values obtained in similar vertical heterostructures \cite{Abanin2011,Ribeiro2018,Li2020,Aharon-Steinberg2021}.\\
Figure \ref{fig:filling} displays the longitudinal resistance and the filling factor as a function of $V_g$ measured at 1.5 K and 12 T. On top of the standard evolution of the Landau levels for single layer graphene $\nu = \pm 2, \pm 6, \dots$ we can clearly observed unusual plateaus of conductance where both spin and valley degree of freedom have been lifted such as $\nu=0, \nu=\pm 1$, and eventually fractional quantum Hall plateaus. No traces of the interference with Landau levels arising from secondary Dirac cones have been found thorough the whole study.

\begin{figure*}[!h]
\includegraphics[width=15cm]{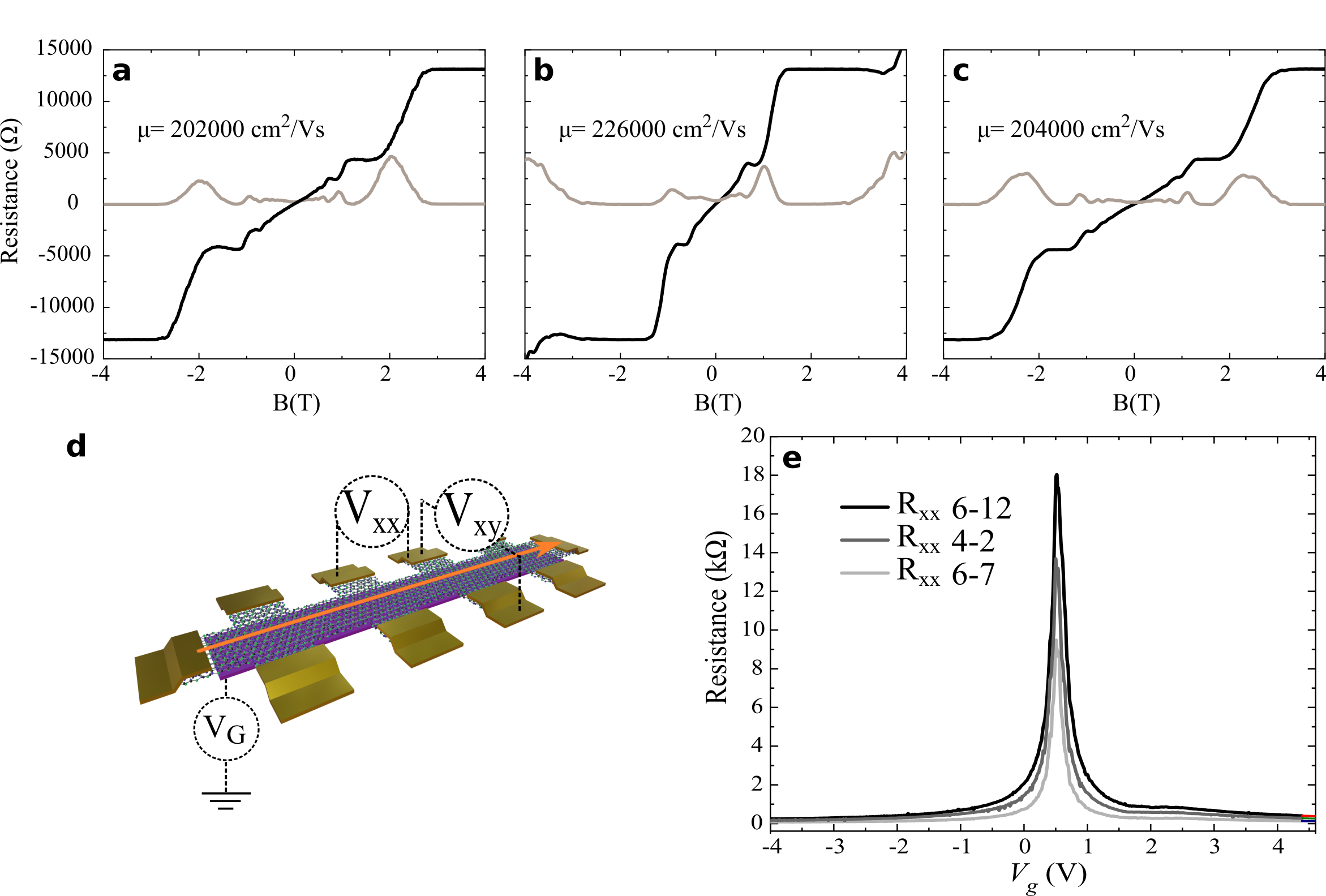}
\caption{\label{fig:mobility}SUPPORTING INFO 5. \textbf{a}, \textbf{b}, \textbf{c}, Spectra for the local longitudinal $R_{\tiny{XX}}$ (in grey) and Hall resistances $R_{\tiny{XY}}$ (black) as a function of the external magnetic field measured at constant gate voltage of $0\,$V (\textbf{a}), $0.3\,$V (\textbf{b}) and $1.5\,$V (\textbf{c}) respectively. Each panel includes the extracted mobility value. \textbf{d}, Sketch of a typical Hall-bar-like configuration of the contacts where the solid line represents the longitudinal glocal driving current. \textbf{e}, Local resistance vs. backgate voltage between three different sets of contacts at B=0T. Charge neutrality point (CNP) appears as a Dirac peak at $V_{G}\sim$0.6V }
\end{figure*}

\begin{figure*}
\includegraphics[width=15cm]{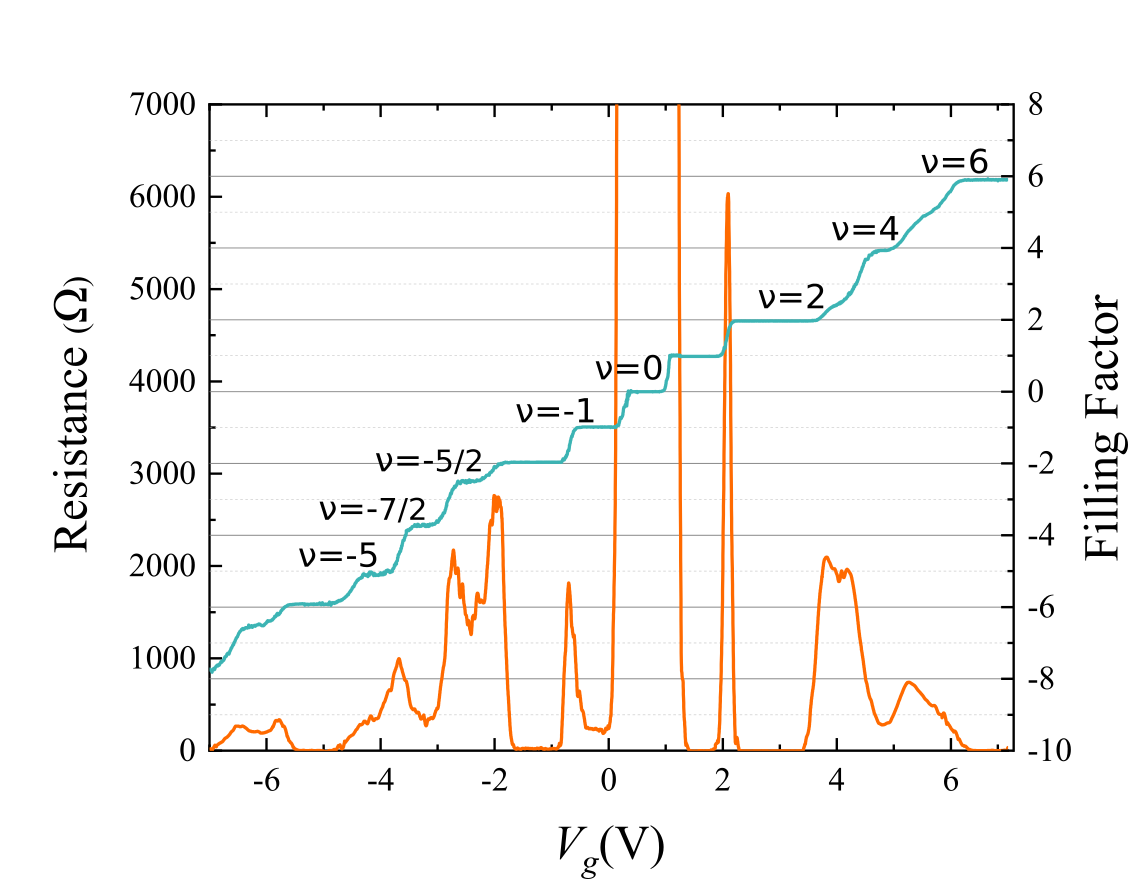}
\caption{\label{fig:filling}SUPPORTING INFO 6. Magnetoresistance (left axis) and filling factor (right axis) versus gate voltage acquired at B = 12 T and T = 1.5 K. Current was injected between contacts 5 and 14, and the Hall and longitudinal resistances were measured between contacts 3 and 7 and 3 and 2, respectively.}
\end{figure*}

\clearpage
\subsection{\label{subsec-Rnl}Evolution of the non-local resistance with magnetic field, distance and temperature.}

The Ohmic contribution to the non-local signal as a function of the distance ($\Delta x$) to the driving current can be extracted from the relation $R_{nl}=V_{nl}/I_0\sim (4\pi)\rho_{xx}e^{-\rvert \Delta x\rvert/\lambda}$ as shown in Refs.~\cite{Abanin2011,Ribeiro2018,Aharon-Steinberg2021} being $\lambda=\text{W}/\pi$, W the bar width and $\rho_{xx}$ the longitudinal conductivity. This relation serves us to extract the expected ratio between the non-local resistances at different distances from a pure Ohmic-bulk contribution. The enormous discrepancy between measured and expected ratios in the non-local signals  $\left(\frac{R_{nl}^{\Delta x_2}}{{R_{nl}^{\Delta x_1}}}\right)_\text{meas}=0.29 >> \left(\frac{R_{nl}^{\Delta x_2}}{{R_{nl}^{\Delta x_1}}}\right)_\text{Ohmic}=0.005 $ rules out trivial Ohmic contributions to $R_{nl}$. Black dashed lines show the expected evolution for $R_{nl}(\Delta x)$ following the Ohmic-contribution-relation.\\

Chiral non-local resistances are robust as a function of the electronic temperature as shown in Fig.\ref{non-local-resistance-vs-T} where $R_{nl}$ and $R_{\text{local}}$ as a function of $V_{g}$ are displayed for different temperatures under external low fields of $B=0.1\,$T. While the maximum value for the different resistances slightly drops with $T$, chirality is clearly preserved. \\

Finally, Figure \ref{normalized-Rnl} shows the evolution of the normalized $R_{nl}$ as a function of $V_g$ measured at different external magnetic fields at a electronic temperature of 250 mK. Every non-local resistance has been normalized by its local counterpart for clarity and they have been recorded following the sketch in the figure. As we can see from both panels the normalized non-local signals preserve a strong chirality and are enhanced by the application of small to moderate external magnetic fields. As the orbital magnetic moment in graphene interacts with the external magnetic field the interplay between the Zeeman shift and the transverse flow of orbital magnetic moments increases giving rise to a strengthened non-local signal.

\begin{figure*}[!h]
\includegraphics[width=15cm]{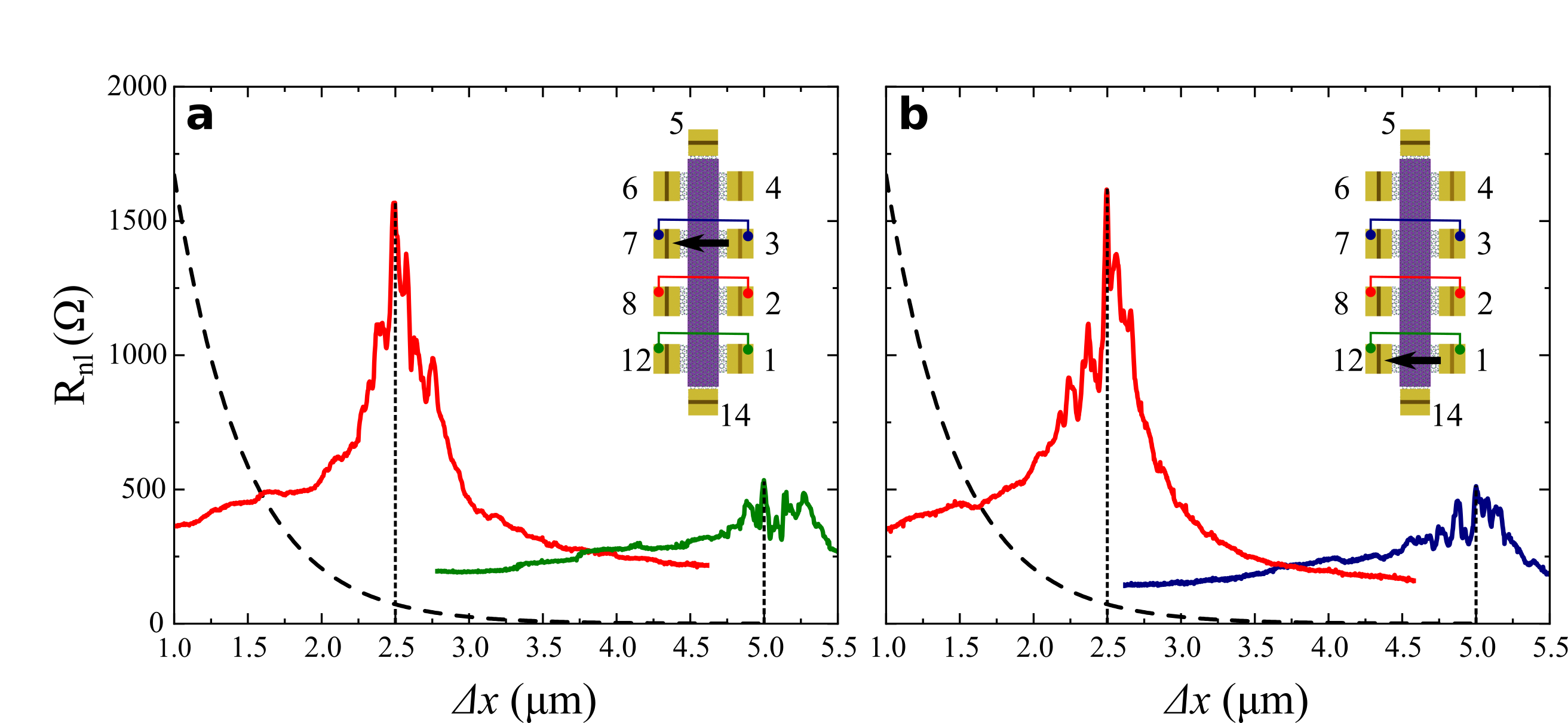}
\caption{\label{exponential-decay}SUPPORTING INFO 7. \textbf{a},\textbf{b}, Non local resistance peaks for two different configurations depicted on the schematic inset for each graph. The Ohmic contribution to the non-local signal has been calculated following the exponential dependence (black dashed line) with the distance $R_{nl}=V_{nl}/I_0\sim (4\pi)\rho_{xx}e^{-\rvert \Delta x\rvert/\lambda}$ , where we have taken the experimental value of $\lambda=W/\pi\,$ being $W=1.5\,\mu$m the bar width and $\rvert x\rvert$ the distance between local and non-local pairs of contacts as found in references \cite{Abanin2011,Ribeiro2018,Aharon-Steinberg2021}.}
\end{figure*}

\begin{figure*}
\includegraphics[width=15cm]{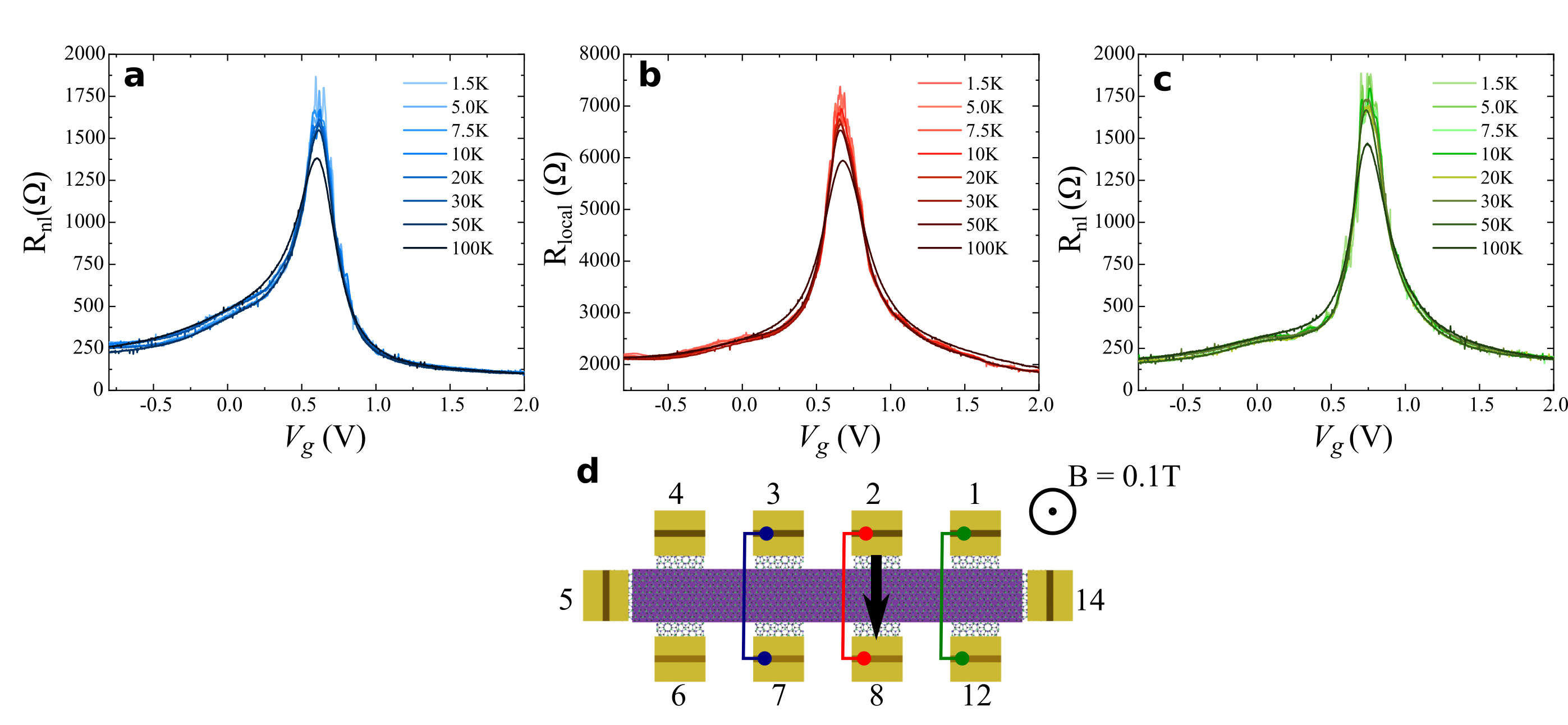}
\caption{\label{non-local-resistance-vs-T}SUPPORTING INFO 8. \textbf{a},\textbf{b},\textbf{c}, Evolution of the local and non-local resistance versus $V_g$ measured at $B=0.1\,$T at different electronic temperatures. A noticeable chirality is preserved up to temperatures $T>100\,$K. \textbf{a},\textbf{c}, display the non-local resistances in a symmetric configuration while (\textbf{b}) shows the local resistance. \textbf{d}, Sketch of the sample where non-local pairs of contacts have been placed symmetrically to the local driving current marked with a solid arrow.}
\end{figure*}

\begin{figure*}
\includegraphics[width=15cm]{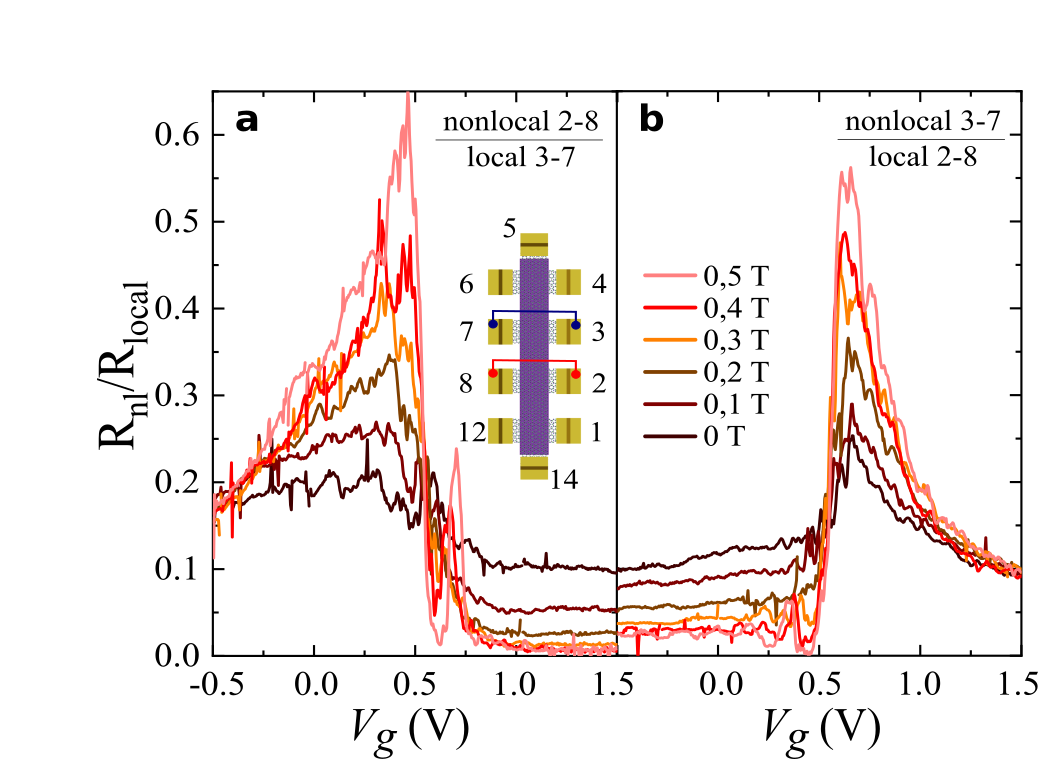}
\caption{\label{normalized-Rnl}SUPPORTING INFO 9. \textbf{a},\textbf{b}, non-local resistances as a function of $V_g$ normalized by their corresponding $R_l$ at different external magnetic fields. All curves were measured at 250 mK in two different local-to-non-local configurations as displayed in the sketched sample. Chirality in the non-local signal becomes more pronounced as the magnetic field increases.}
\end{figure*}

\clearpage
\subsection{Ohmic and thermal contribution to $R_{nl}$}

Ohmic (thermal) contribution is described by the van der Pauw formula, 

\begin{equation}
    R_{nl,\Omega}=\frac{\text{W}}{\pi \text{L}}R_{xx}\,ln\left[ \frac{\cosh{(\pi \text{L}/\text{W})}+1}{\cosh{(\pi \text{L}/\text{W})} - 1} \right]
\end{equation}

where L and W are the channel length and width. In zero magnetic field and for a maximal value of $R_{xx}=18\,$k$\Omega$ and being L/W$=2.5/1.5$ we would obtain $R_{nl,\Omega}=73\,\Omega$ which is two orders of magnitude smaller than the measured $R_{nl}$. Moreover, in absence of external magnetic field only Joule heating effect contributes to the second harmonic of non-local signal $R^{2f}$. We were unable to measure significant $R_{nl}^{2f}$ at zero magnetic field field while for $B=500\,$mT we obtained a signal smaller than $2\%$ as shown of $R_{nl}$ as shown in Supplementary \ref{fig:2ndharmonic} where $R_{nl}^{1f}$ and the envelope signal of $R_{nl}^{2f}$ multiplied for a prefactor $\times 50$ are displayed fro two different pair of non-local contacts. 

\begin{figure*}[!h]
\includegraphics[width=15cm]{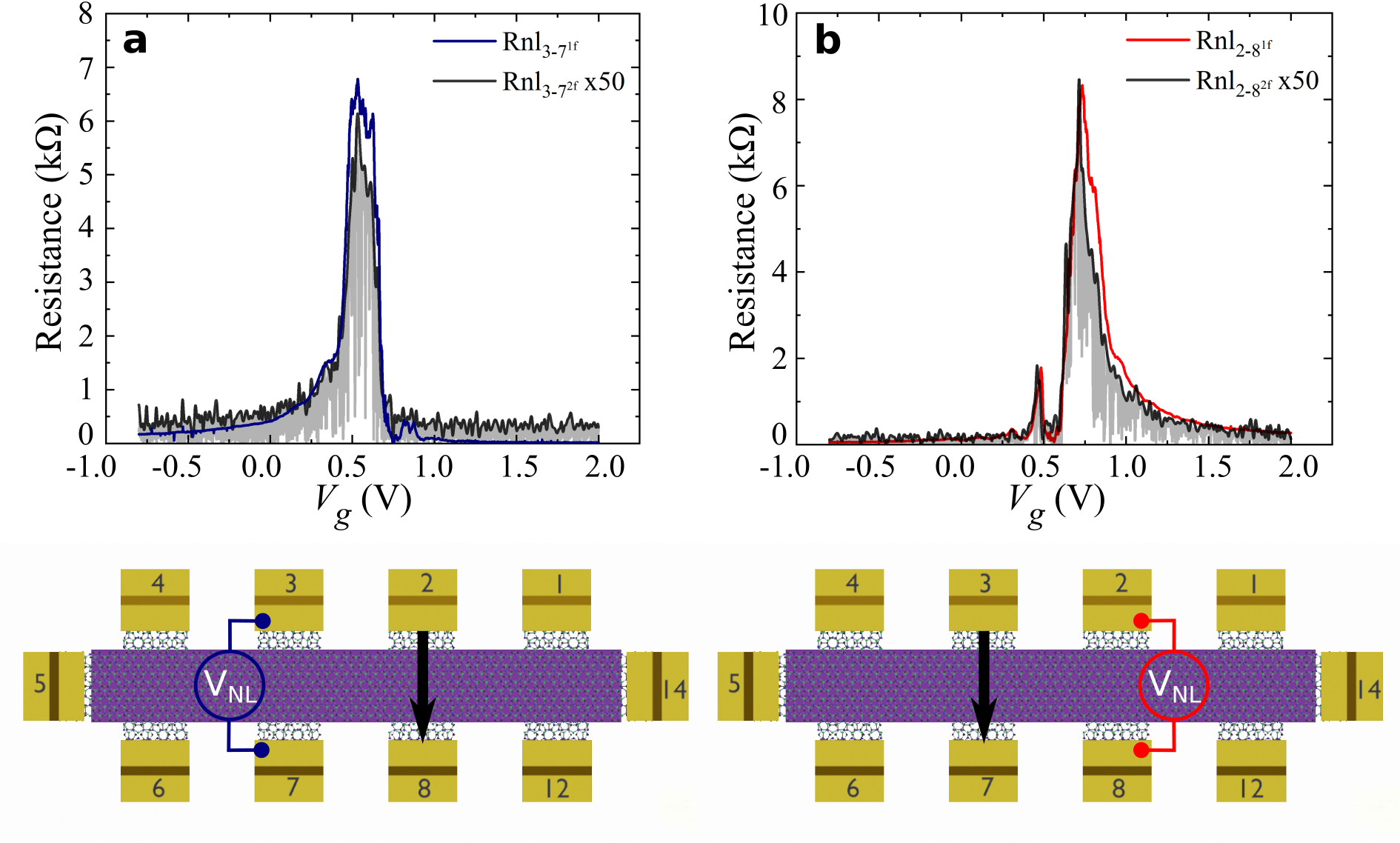}
\caption{\label{fig:2ndharmonic}SUPPORTING INFO 10. Comparison of the non-local resistance and second harmonic signals for two different configurations as indicated in the bottom sketchs. Data were recorded at $1.5\,$K and with an external applied magnetic field $B=500\,$mT. The intensity of the second harmonic has been adjusted in order to verify the match between the signals, which indicates a negligible influence of thermal phenomena to which the second harmonic would be sensitive, including Nernst and Ettinghsausen effects.}
\end{figure*}

\subsection{Perturbation theory and Linear response formalism for the orbital Hall effect \label{supp:OHE}}

Here, we present a general discussion on perturbation theory to include the effect of magnetic field up to the first-order in the orbital Hall effect calculation. We assume that the intensity of the magnetic field is weak enough and the system is far from the Landau level regime. In this situation, we can treat the effect of the magnetic field in the framework of perturbation theory following Ref. \cite{Bloch-Orbital-Moment-Khon-1, Magnetic-Field-perturbation-theory-1, Magnetic-Field-perturbation-theory-2}. In graphene systems, this situation is observed in experiments for the magnetic field with intensities $|\vec{B}|\lesssim 1.0 T$. We also present the linear response formulas for Hall conductivity (HC) and orbital Hall conductivity (OHC) used to obtain the theoretical contour plots  presented in the main text. 

Despite the complexity of graphene/$h$-BN heterostructure used in experiments, near the Dirac point, it is possible to analyze the physics of the system using simply the Hamiltonian of graphene monolayer (ML) with mass \cite{Gorbachev2014, Graphene-HBN-Miller, VHE-Experiment-Old-Interpretation, Song2015}. This simplification should not be applied to secondary peaks of non-local resistance measurements, but may be used to understand some features of the central peak. Writing the Hamiltonian of graphene monolayer (ML) on the tight-binding basis, $\beta^{ML}_{tb}=\{ A, B\}$, and expanding it near the valleys $\vec{K}=\left( 4\pi \right)/ \left( 3\sqrt{3}a\right)\hat{x}$ and $\vec{K}'=-\vec{K}$, we obtain:
\begin{eqnarray}
\mathcal{H}^{ML}=\begin{bmatrix}
   \frac{\Delta}{2} & \gamma_- \\
    \gamma_+ & -\frac{\Delta}{2} 
\end{bmatrix},
\label{HGGM}
\end{eqnarray}
where, $\gamma_{\pm}=\hbar v (\tau q_x \pm i q_y )$ with, $\vec{q}=\vec{k}-\tau \vec{K}$ is the wave vector relative to valleys and, $\tau=\pm 1$ for Dirac cone at valleys $K$ and $K'$ respectively. The velocity $v=3at/2\hbar$, where $t=2.8 \text{eV}$ is the nearest-neighbor hopping amplitude of graphene and $a= 1.42\,$\AA\ is the carbon-carbon distance. $\Delta$ is the mass gap term induced by the $h$-BN substrate. The mass breaks the spatial-inversion symmetry of the graphene monolayer and opens a bandgap with amplitude $E_g=\Delta$ in the electronic spectrum. 

 $E^{v(c),0}_{n, \vec{k}}$ and $\big| u^{v(c),0}_{n,\vec{k}}\rangle$ are the dispersion energies and eigenstates of these unperturbed Hamiltonian. For ML we have $n=v (c)$ to index the state of valence (conduction) band and, for BL we have $n=1 v(c)$, $2 v(c)$ to index the states of the 2-dimensional subspace of the valence (conduction) band. We use these unperturbed energies and states to compute the orbital magnetic moment. Considering the application of a weak magnetic field, the energy of electronic bands is corrected by
\begin{eqnarray}
E_{n, \vec{k}}= E^{0}_{n, \vec{k}} - \langle u^{0}_{n, \vec{k}} \big| \Big(\hat{\vec{m}}_{\vec{k}} \cdot \vec{B} \Big) \big| u^{0}_{n, \vec{k}} \rangle, \label{EB}
\end{eqnarray} 
where the second term on the right-hand side of the equation is the orbital magnetic moment $m_{n,n}^z(\vec{k})$ of $n$-th electronic states. The correction in electronic states is
\begin{eqnarray}
\big| u_{n, \vec{k}}\rangle = && \mathcal{N}\Bigg[ \big| u^{0}_{n, \vec{k}}\rangle \nonumber \\
&&- \sum_{m \neq n} \frac{\langle u^{0}_{m, \vec{k}} \big| \Big( \hat{\vec{m}}_{\vec{k}} \cdot \vec{B} \Big) \big| u^{0}_{n, \vec{k}} \rangle}{\big( E^{0}_{n,\vec{k}} - E^{0}_{m, \vec{k}}\big)} \big| u^{0}_{m, \vec{k}}\rangle \Bigg]. 
\label{uB}
\end{eqnarray}
In Eq. (\ref{uB}), $\mathcal{N}\big[ . . . \big]$ means that, it is necessary to normalize the state inside the square brackets after including the perturbative contribution. The equations above give the first-order perturbation theory in the linear order of the magnetic field. In the case of graphene ML, the correction in Eq. (\ref{uB}) vanishes due to the dimensionality of Hilbert space, i.e., one-dimensional Hilbert space on $c(v)$-bands. 
 
\begin{figure}[h!]
	\centering
	 \includegraphics[width=0.65\linewidth,clip]{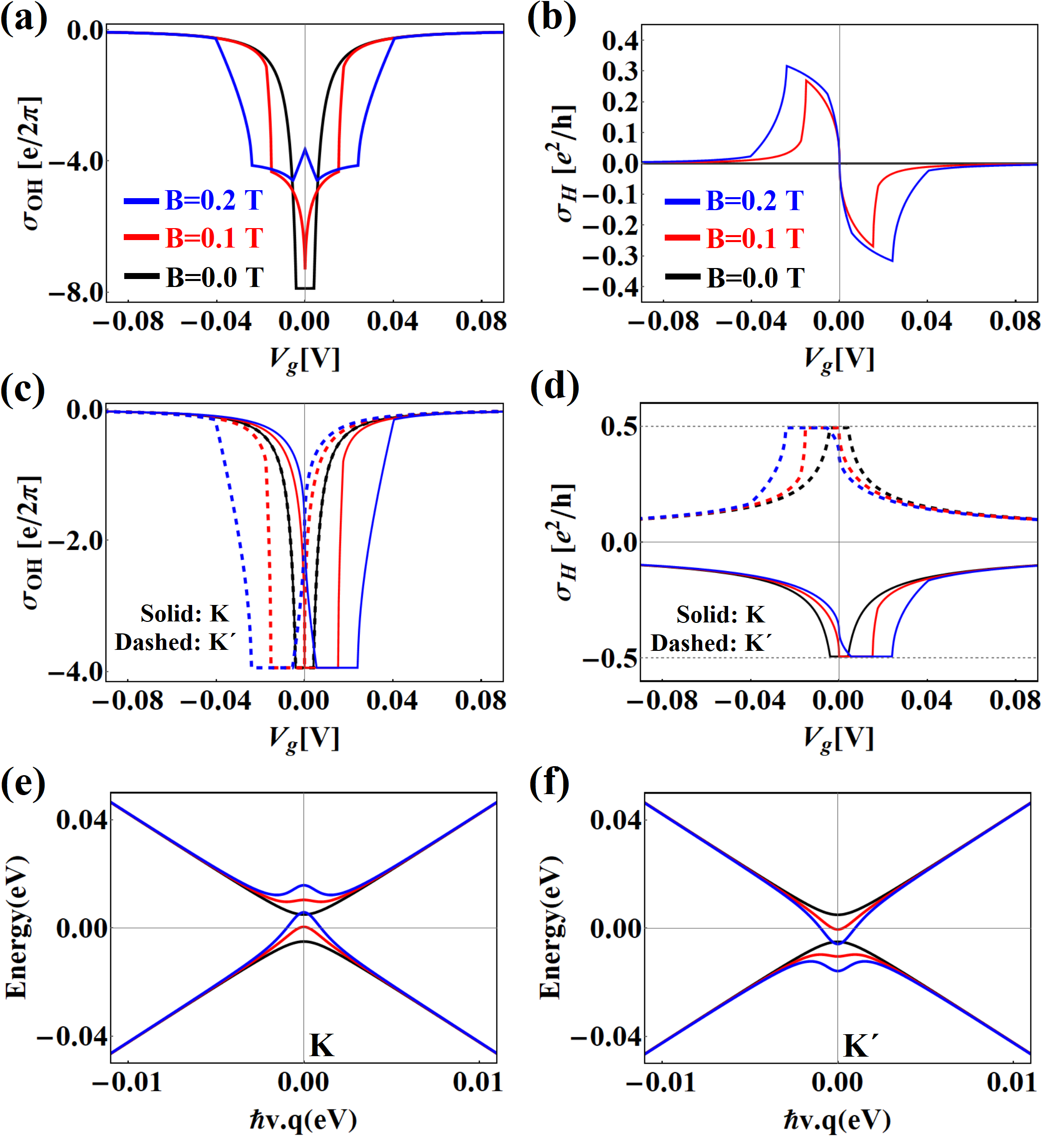}   
	\caption{SUPPORTING INFO 11. Orbital Hall conductivity \textbf{a} and Hall conductivity \textbf{b} of graphene ML for three values of magnetic fields, $B=0.0$ (black), $0.1$ (red), and $0.2$ (blue) T.  Panels \textbf{c} and \textbf{d} show the contributions for conductivities of different valleys $K$ (solid) and $K'$ (dashed). Here we used an energy gap parameter $\Delta= 10 \text{meV}$.}
	\label{fig:figM2}
\end{figure}

With the corrected energies and states of Eqs. (\ref{EB}, \ref{uB}), we compute the HC and OHC. In the low-temperature limit, the HC is
\begin{eqnarray}
\sigma_H (E_f)=\frac{e^2}{\hbar}\sum_{n} \int \frac{d^2k}{(2\pi)^2}\Omega_{n}(\vec{k})\Theta \big(E_f-E_{n, \vec{k}} \big),  \ \ \label{sigmaH}
\end{eqnarray}
\begin{eqnarray}
\frac{\Omega_{n}(\vec{k})}{2\hbar^2}=\sum_{m\neq n} \text{Im} \left[ \frac{\langle u_{n, \vec{k}}\big|\hat{v}_x(\vec{k})\big| u_{m, \vec{k}}\rangle \langle u_{m, \vec{k}}\big|\hat{v}_y(\vec{k})\big| u_{n, \vec{k}}\rangle}{\Big(E_{n, \vec{k}}-E_{m, \vec{k}} \Big)^2} \right]. \nonumber \\
\label{OmegaH}
\end{eqnarray}
The OHC is given by
\begin{eqnarray}
\sigma_{OH} (E_f)=e\sum_{n} \int \frac{d^2k}{(2\pi)^2}\Omega^{OH}_{n}(\vec{k})\Theta \big(E_f-E_{n, \vec{k}} \big), \nonumber \\
\label{sigmaOH}
\end{eqnarray}
\begin{eqnarray}
\frac{\Omega^{OH}_{n}(\vec{k})}{2\hbar}=\sum_{m\neq n} \text{Im} \left[ \frac{\langle u_{n, \vec{k}}\big|\hat{v}_x(\vec{k})\big| u_{m, \vec{k}}\rangle \langle u_{m, \vec{k}}\big|\hat{j}^{L_z}_y(\vec{k})\big| u_{n, \vec{k}}\rangle}{\Big(E_{n, \vec{k}}-E_{m, \vec{k}} \Big)^2} \right]. \nonumber \\
\label{OmegaOH}
\end{eqnarray}

In Eqs. (\ref{OmegaH}, \ref{OmegaOH}), the velocity operators are defined by, $\hat{v}_{x(y)}(\vec{k})=\hbar^{-1}\partial \mathcal{H}(\vec{k})/ \partial k_{x(y)}$. In Eq. (\ref{OmegaOH}), the current with OAM polarized in z-direction that flows in the y-direction is $\hat{j}^{L_z}_y(\vec{k})=\left(L_z(\vec{k}) \hat{v}_y(\vec{k})+ \hat{v}_y(\vec{k})L_z(\vec{k}) \right)/2$, with, $L_z(\vec{k})=-(\hbar/\mu_B)\hat{m}^z_{\vec{k}}$. $\mu_B=(e\hbar)/(2m_e)$ is the atomic Bohr magneton defined using the electron rest mass $m_e$.

\subsection{Technical details of the non-local Resistance Simulations}

In what follows, we give a brief overview of the details of the non-local resistance simulations. As mentioned in the main text, Marmolejo-Tejada \textit{et al.} demonstrated that the simplistic model that considers only $p_z$ electrons in graphene does not reproduce the electronic structure of zigzag Gr/hBN nanoribbons \cite{marmolejo2018deciphering}. These results motivated us to construct a more involved tight-binding model to provide an appropriate description of the energy states of these nanoribbons. Based on previous first-principles works \cite{TBtheoryGr}, we took advantage of the $D_{3h}$ symmetry of the $\pi$-bands of graphene near the Brillouin zone corners and constructed a model composed of the $p_z$ and the $d_{xz},d_{yz}$ orbitals. Using the Slater-Koster \cite{SlaterKoster} parameters of reference \cite{boykin2011accurate}, we built a $6$-bands tight-binding model that considers only nearest-neighbour hopping integrals. Figure \ref{fig:SupModelCompare} portrays the energy states of nanoribbons of $50$ nm breadth for the $p_z$ model and the $6$-bands tight-binding model. Upon brief inspection, it is clear that the simplistic $p_z$ model, though capable of capturing most of the bulk properties of monolayer graphene and other heterostructures, misrepresents the behaviour of the edge states leading to the appearance of an energy gap when the staggered sublattice potential breaks the inversion symmetry. In contrast, the $6$-bands tight-binding model allows the hybridization between $p_z$ and $d_{xz},d_{yz}$ orbitals, promoting the appearance of massive gapless edge-states that gives rise to the non-local signals in our simulations in agreement with the results from Marmolejo-Tejada \textit{et al.} \cite{marmolejo2018deciphering} and supports the formation of orbital magnetic moments that can couple with the external magnetic field. Additionally, the $6$-bands model reproduces the width-independent insulating behaviour of armchair graphene nanoribbons reported in Ref. \cite{ArmchairGapsSon}.
The model up to 5 nearest-neighbours exposed in Ref. \cite{marmolejo2018deciphering} has been tested as well in an equivalent device, finding non-local resistances of the same order of magnitude of Fig. \ref{fig:nlfig}. However, the profile of the non-local resistance does not resemble the experimental anisotropic peak and the interpretation up to long distance neighbours is not as clear as the multi-orbital model. \\

\begin{figure*}
\includegraphics[scale=0.15]{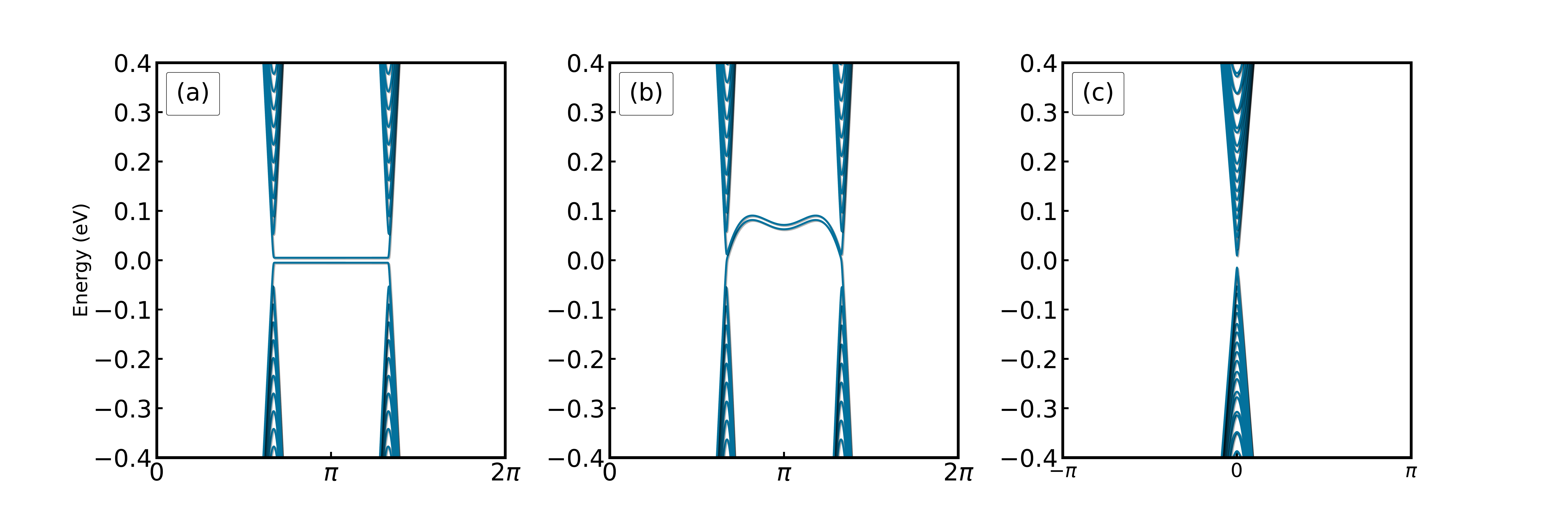}
\caption{\label{fig:SupModelCompare} SUPPORTING INFO 12. Comparison of the energy bands of zigzag Gr/hBN nanoribbons with $50$ nm breadth and staggered sublattice potential $\Delta=5\text{ meV}$ for: \textbf{a} model with $p_z$ orbitals, and \textbf{b} model with $p_z$,$d_{xz}$ and $d_{yz}$ orbitals. \textbf{c} Energy bands of Armchair Gr/hBN nanoribbons with the same width and staggered sublattice potential. }
\end{figure*}

In quantum transport simulations, we used an $8$-terminal geometry depicted in the inset of panel \textbf{c} of figure \ref{fig:nlfig} in which we fixed the width of the channel and the leads to $13$ nm and the separation between each pair of leads perpendicular to the channel direction to $157$ nm. Besides this, we also doped the leads perpendicular to the ribbon to counter the insulating nature of the armchair Gr nanoribbons and avoid contact resistance in our simulations. We computed the non-local resistance by obtaining the transmission probabilities for all the contacts and constructed the conductance matrix. Then, we determined the voltages in the leads using the equation $I=GV$, and to ensure the charge conservation within the system fixed the voltage of the lead from $2$ to zero. To include the effects of magnetic fields in our simulation, for the scattering region we have redefined the hopping integrals like ${t_{ij}}^{\mu \nu} \rightarrow {t_{ij}}^{\mu \nu}\times\exp{\left(-i\frac{e}{\hbar}\int_{\vec{R}_i - \vec{R}_j}\vec{A}\cdot d\vec{l}\right)}$, where $\vec{A} = -By\hat{x}$ is the vector potential associated to the field chosen to not depend on the periodic direction of the ribbon.\\

\begin{figure*}
\includegraphics[scale=0.20]{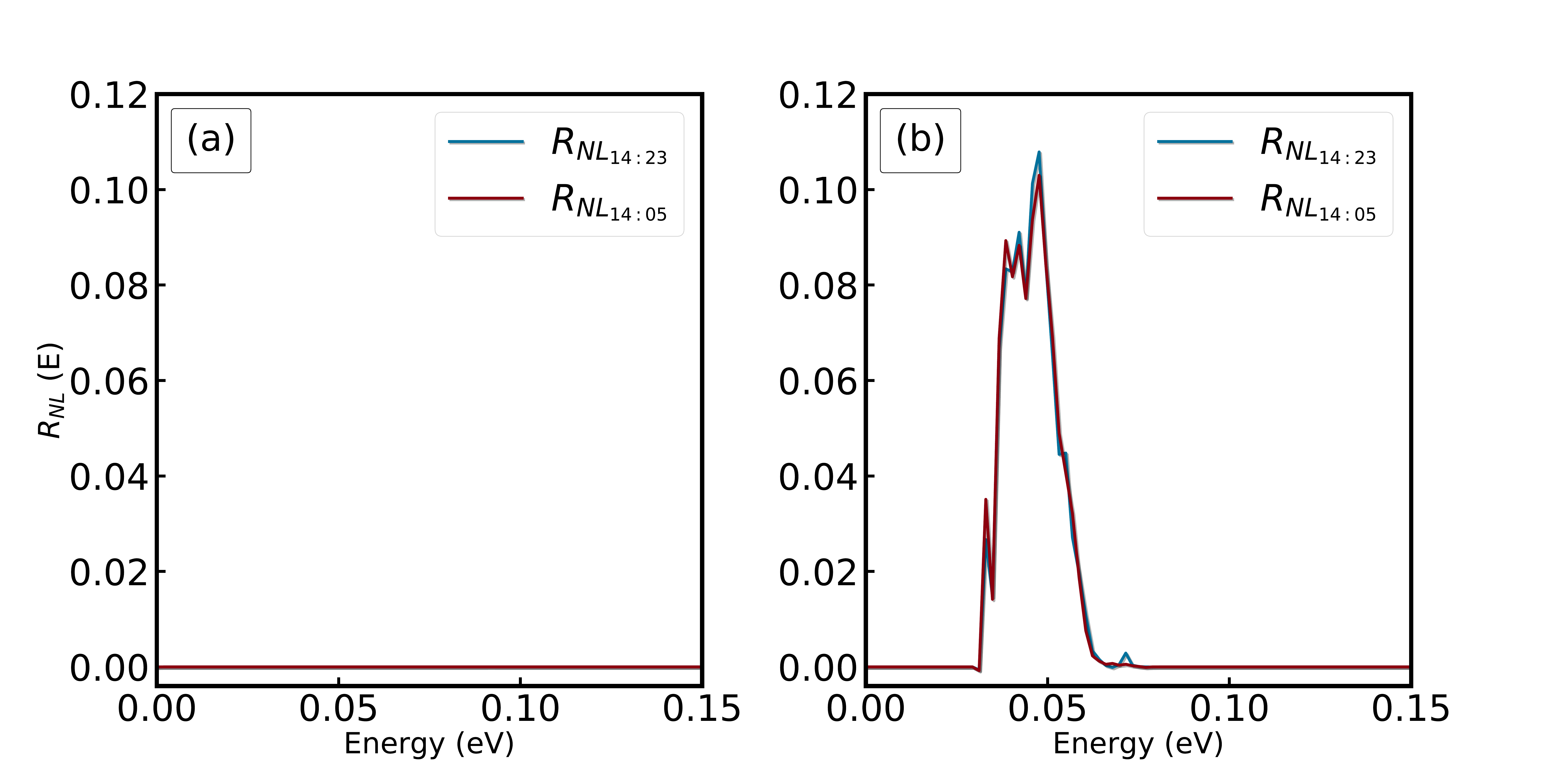}
\caption{\label{fig:SupModelCompareNL} SUPPORTING INFO 13. non-local resistance for current injection across $1 - 4$ in $8$-terminal devices with $13\,$nm breadth and $157\,$nm distance between the injection and collection leads for: \textbf{a} the model with $p_z$ orbitals and \textbf{b} the $6$-bands model.
}
\end{figure*}

\end{widetext}

\end{document}